\documentclass[aps,twocolumn, nofootinbib, superscriptaddress]{revtex4}

\usepackage{graphicx}
\usepackage{verbatim}           
\usepackage{simplewick}
\usepackage{amsmath}
\usepackage{array}
\usepackage{bbold}

\newcommand{\bea}{\begin{eqnarray}}
\newcommand{\eea}{\end{eqnarray}}
\newcommand{\be}{\begin{eqnarray}}
\newcommand{\ee}{\end{eqnarray}}
\newcommand{\nn }{\nonumber}

\newcommand{\D}{\mathcal{D}}

\newcommand{\m}{{\bf m}}
\newcommand{\n}{{\bf n}}
\newcommand{\x}{{\bf x}}
\newcommand{\y}{{\bf y}}
\newcommand{\z}{{\bf z}}
\newcommand{\kk}{{\bf k}}

\begin{document}

\title{Long-Distance Quantum Transport  Dynamics in Macromolecules } 
\author{E. Schneider}
\email{schneider@science.unitn.it}
\affiliation{Physics Department, Universit\`a degli Studi di Trento, Via Sommarive 14, Povo Trento}
\affiliation{Trento Institute for Fundamental  Physics and Applications (TIFPA), Via Sommarive 14, Povo Trento} 
\author{P. Faccioli}
\email{faccioli@science.unitn.it}
\affiliation{Physics Department, Universit\`a degli Studi di Trento, Via Sommarive 14, Povo Trento}
\affiliation{Trento Institute for Fundamental  Physics and Applications (TIFPA), Via Sommarive 14, Povo Trento} 

\begin{abstract}
Using renormalization group methods, we develop a rigorous coarse-grained representation of the  dissipative dynamics of quantum excitations propagating inside open macromolecular systems. We show that, at very low spatial resolution, this quantum transport theory reduces to a modified Brownian process, in which quantum delocalization effects are accounted for by means of an  effective  term in the Onsager-Machlup functional. Using this formulation, we derive a simple analytic solution for the time-dependent probability of observing the quantum excitation at a given point in the macromolecule. This formula can be used to predict the migration of natural or charged quantum excitations in a variety of molecular systems including biological and organic polymers,  organic crystalline transistors or photosynthetic complexes. For illustration purpose, we apply this method to investigate intelastic electornic hole transport in a long homo-DNA chain.
\end{abstract}
\maketitle

\section{Introduction}

The striking observation of long-lived coherent quantum energy transport in photosynthetic systems~\cite{photo_exp1}, and the perspective of realizing nano-scale molecular devices with specific opto-electronics properties~\cite{nano-bio-electronics} has motivated an increasing effort towards investigating the propagation of charged and neutral quantum excitations across many organic~\cite{NEGF1, organic1, organic2, organic3, P3HT1} and biological \cite{DNAreview, DNA_theory1, DNA_theory2, DNA_theory3, peptide,  photo_theory1,  photo_theory2, plenio2, plenio3, photo_theory3, photo_theory4} macromolecules.

In contrast to the electric conduction in metals, quantum transport in soft condensed matter can be significantly  influenced by the coupling  to the molecular vibrations  and to the surrounding environment. Consequently, the natural theoretical framework to describe the propagation of excitons, electrons and holes  through macromolecules is that of the open quantum systems~\cite{open_quantum_system_book}.  
 
In this context,  theoretical models have been developed
in which the dynamics of the quantum excitation is coarse-grained at the level of a simple one-body Hamiltonian, while the coupling to the molecular motion and the environment  is collectively represented by means of an effective bosonic bath  (see e.g. Ref.~\cite{DNA_theory1,photo_theory1,photo_theory4} ). 
These  models provide conceptually sound and computationally efficient tools to investigate the \emph{general} mechanisms which underlie the long-range charge transport and the loss of quantum coherence in macromolecules. On the other hand,  the lack of chemical detail makes it difficult to obtain quantitative predictions on the quantum transport properties in \emph{specific} molecular systems. 

Complementary theoretical approaches have been developed which encode much information about the specific chemical structure of the macromolecule, hence are in principle better suited to obtain quantitative predictions. These models are generally based on combining the Schr\"odinger equation for the one-body wave function of the  quantum excitation with molecular dynamics (MD) simulations for the motion of the atomic coordinates (see e.g. \cite{P3HT1, DNA_theory3, boninsegna}). The quantum excitation's dynamics in  these models can be investigated in great detail through extensive numerical simulations.  On the other hand, the lack of analytic insight makes it difficult to identify the physical mechanisms which are responsible for the transport dynamics. 

In a recent paper~\cite{Elia1}, we developed a microscopic theoretical framework to describe quantum transport in macromolecules, which combines  chemical  detail  with analytic insight.  This  approach is based on a coherent quantum field path integral representation of the system's reduced density matrix.  
The path integral formalism is convenient because it allows to rigorously trace out from the density matrix the atomic coordinates, hence avoids computationally expensive MD simulations.  
The quantum field theory (QFT) formalism is adopted because it drastically simplifies the description of the dynamics of  open quantum systems.
Indeed, we have shown that, using QFT,  the problem of computing real-time evolution of observables in an open system can be mapped into the problem of computing vacuum-to-vacuum Green's functions in some virtual system. 

In the short-time limit, such Green's function can be computed in perturbation theory, using standard Feynman diagram techniques based on time-ordered
propagators (rather than time-directed propagator), without having to perform any numerical MD simulation. This feature makes the calculation of the density matrix
computationally inexpensive or, in some cases, even analytic. In addition, the diagrammatic expansion offers physical insights, e.g. about the specific processes
which are responsible for quantum decoherence and dissipation.

Unfortunately, such a perturbative approach  breaks down in the long-time regime, when the propagation is dominated by multiple scattering processes and typically becomes inapplicable after $50-100$ fs. Hence, the investigation of quantum transport over  larger distances and longer time intervals using the QFT formalism requires a fully non-perturbative approach. 

In this work we  tackle this problem using the renormalization group (RG) formalism to systematically coarse-grain the dynamics. The result is a rigorous ``low-energy"  approximation of our original microscopic QFT which by yields the same dynamics in the limit in which the quantum excitation travels for a long-time and covers distances which are large compared to its De Broglie's thermal wavelength. 

The advantage coarse-graining is that the resulting ``low-resolution" effective field theory (EFT) is much simpler than the corresponding microscopic QFT and admits an analytical solution. In particular, we shall show that the probability density $P(\y, t |\x)$ for a quantum excitation initially produced at the point $\x$ to be found at the $\y$ after a time $t$ can be written in the path integral form 
\be\label{PI}
&& \hspace{-3mm} P({\bf y}, t| {\bf x},0) = \int_{\bf x}^{\bf y} \hspace{-0.4mm} \mathcal{D}{\bf R}~e^{-\int_0^t d\tau \left[\frac{\dot{\bf R}^2}{ 4 D_0 } -\xi^2   (~C_2~\dot{\bf R}^2 - C_4 \dot{\bf R}^4) +\ldots\right]}. \nn\\
\ee
The inner products which appear at the exponent in this equation are defined by a metric tensor which is in general non-diagonal, while  $\xi^2\propto \hbar^2$ is a small expansion parameter to be defined below and the dots represent terms which are irrelevant in the large-distance and long-time  limit. Hence, the propagation of the quantum excitation corresponds to a modified anisotropic diffusion process, in which the quantum effects are by a set of  effective parameters and operators.

The effective parameters $D_0, C_2, C_4$, as long as the entries of the metric tensor defining the inner product encode the information about the unresolved short-distance physics, such as the molecule's electronic structure and vibrational spectrum. All these parameters can be determined  by measuring the quantum excitation's mobility. 

In the physical regime  in which the path integral Eq. (\ref{PI}) holds it is even possible to derive a simple analytic expression for its solution, by exploiting the fact that $\xi\ll 1$. Such a solution offers a powerful tool  to investigate complex non-equilibrium quantum transport processes, within a very simple framework.

The paper is organized as follows. In the next section we review the EFT formalism, which allows to construct  rigorous  low-energy approximations to physical theories. In sections \ref{model} and \ref{RQFT}, we first define the Hamiltonian of a microscopic model for quantum transport and then review its equivalent QFT formulation, that can be used to compute perturbatively the time-dependent density-matrix, in the short-time regime. In section \ref{long-time}, we use the EFT framework to derive a systematic approximation of such a QFT, which yields the same dynamics in the large-distance and long-time regimes. Next, in sections \ref{effstoch} and \ref{QEFTfunc}, we analyze the long-time regime and we show that our EFT can be formulated as a modified diffusion process. In section \ref{solutionandrenormalization} we derive the analytic expression for the probability density and discuss the renormalization of the EFT.  Section \ref{application} is devoted to the illustrative application of this framework to 
investigate hole propagation in homo-DNA. Finally, conclusions and outlooks are summarized in section \ref{conclusions}.  

\section{The effective field theory formalism}
\label{EFT}

The internal dynamics of macromolecules is characterized by a multitude of time-scales, which are spread over many orders of magnitude. Conformational transitions typically occur beyond the ns time scale, hence are clearly decoupled from the local   thermal vibrations and the solvent-induced dissipative relaxation times, which occur at the ps scale.  The hopping of quantum excitations across nearby molecular molecular orbitals and the loss of quantum coherence crucially depend on the specific chemical structure of the molecule, but typically range from a few fs to fractions of ps. 

The existence of relatively large gaps between the different characteristic time scales in macromolecules naturally suggests to apply RG methods to device rigorous low-energy descriptions of the  dynamics.   
In particular, the EFT formalism (for a pedagogical introduction see e.g. Ref.s \cite{EFT1, EFT2} ) provides a very practical implementation of the RG scheme, which is both rigorous and systematically improvable. Despite these features, to date there have been only a few applications of this framework to quantum transport problems~\cite{graphene1, graphene2} and to   conformational dynamics in macromolecules~\cite{RGMD1, RGMD2, RGMD3}. 

The main idea underlying the EFT approach is the familiar observation that an experimental probe with a given wave-length $\lambda$ is insensitive to the  \emph{details} of the  physics associated to length scales $ \ll \lambda$ and time intervals $\ll \lambda/v$ (where $v$ is the velocity of  propagation of the probe). This fact can be exploited to build in a rigorous way predictive physical theories in which only the degrees of freedom that can be resolved by the probe's wavelength are treated explicitly, while all the ultra-violet (UV) effects (i.e. the physics which is not resolved by the probe) are treated at the implicit level, through a set of local interactions and effective parameters. 

A familiar example of this type of approach is the multi-pole expansion in classical electrodynamics: the soft components of the classical radiation generated by an arbitrarily complicated current source ${\bf J}({\bf r}, t)$ of size $d$ can be replaced by the radiation generated by a sum of point-like multipole currents $E1,M1,\ldots$. In such an expansion, the multipole coefficients implicitly account for the UV physics, which is associated to short distances, of the order of $d$. Only a finite number of multipole terms are  needed to reproduce to any arbitrary (but \emph{finite}) level of accuracy the electro-magnetic radiation at distances $\gg d$. In the following, we shall refer to the physics at length scales much larger than the UV cutoff length-scale as the infra-red (IR) sector of the dynamics.

In the context of  quantum theories,  the EFT scheme is implemented in four steps. First, one introduces the cut-off scale $\lambda$ which defines the separation between the IR physics one is interested in, and the UV physics to be treated implicitly. Next, the most general possible description of the IR dynamics is derived by analyzing the structure and symmetries of the underlying (more) microscopic theory. Then,  a so-called power-counting scheme is introduced in order to identify which coupling terms in the effective theory dominate in the IR limit. Typically, in the QFT formalism, the last step leads to retaining only operators with the lowest number of time and space derivatives of the fields  or of the wave-function.  Finally, through the renormalization procedure, the effective coefficients are determined by matching against experiment or more microscopic calculations, and the dependence on the cut-off is replaced by a (typically much weaker) dependence on the renormalization scale. 

It is important to emphasize that  in EFTs there are no UV divergences, because the cut-off specifies the level of resolution of the theory, hence is kept finite at all times. However, the short-distance physics which is excluded by the finite cut-off is not simply neglected. Instead,  it is accounted for by means of local effective vertexes, called the counter-terms, which enter in the low-energy effective action.  These vertexes parametrize the effects of very short-distance interactions on the long-distance dynamics.

The effectiveness of the EFT scheme depends crucially on the size of the gap separating the IR  and the UV physics. The smaller is the gap, the larger is the number of effective interactions and parameters that have to be introduced in the EFT to reach the desired accuracy. In the absence of a decoupling between IR and UV scales, the dynamics of the EFT depends on  infinitely many effective interactions and parameters, hence looses its predictive power. 

In this work, we exploit the  separations in the length and time scales characterizing the internal dynamics of macromolecular systems to build an EFT for  dissipative quantum transport  in the large-distance and long-time regime. In particular, we restrict our attention to systems in which quantum excitations can propagate over distances much larger than the size  of the individual molecular orbitals and we consider times interval much longer than those characterizing the damping of local thermal conformational oscillations of the macromolecule. Hence, our typical UV cutoff length scale  is of the order of the nm 
and the typical UV time scale is of the order of a few fractions of ps. 

In the next two sections we  begin by reviewing the microscopic field theory description quantum transport we have introduced in Ref.~\cite{Elia1}. 
 Next,  we shall apply the EFT framework to construct a rigorous low-energy approximation of this theory, and then discuss its implications in different time regimes.  
 
 Throughout this work,  we shall adopt a bosonic description of the quantum excitation degrees of freedom, which is clearly appropriate for exciton propagation. However, in the physical limits in which  quantum 
 statistics effects become irrelevant (e.g. in the low-density limit),  the same model can also be used to describe the propagation of fermionic excitations such as electrons or holes. 

\section{A Microscopic Model for Quantum Transport In Macromolecules}
\label{model}

An approach which has been often adopted to model the propagation of quantum excitations within open macromolecular systems (see e.g. \cite{P3HT1, DNA_theory3, boninsegna}) relies on two main approximations: (i) the electronic problem is coarse-grained at the tight-binding approximation level, and (ii)  the dynamics in the absence of quantum excitations is described by the lowest Born-Oppenheimer energy surface. 

The starting point of such  theories is a quantum Hamiltonian which consists of several parts: \be\label{Htot}
\hat H= \hat H_{MC} + \hat H_{M} + \hat H_{CL}.\ee
The quantum transport is pictured as the hopping across different fractional molecular orbitals and follows from the second-quantized Hamiltonian:
\be
\hat H_{MC} = f_{\m \n}[Q]~\hat a^\dagger_{\m} \hat a_{\n}.
\ee
 In this definition the discrete vectors $\m$, $\n$ label the different fractional molecular orbitals (throughout the paper, we shall adopt Einstein's notation, implicitly assuming the summation over repeated  indexes).  $Q=(q_1, \ldots, q_{3N})$ is a vector in the molecule's configuration space (i.e. the set of all the 3N atomic coordinates). In the adiabatic limit,  the matrix elements $f_{\m \n}$ depend parametrically on the molecular orbital wave-functions and on the electronic Hamiltonian $\hat H_{el}$ through the transfer integrals and on-site energies, i.e.
 \be
f_{\bf l m}(Q) =  \langle \Phi_{\bf l} | \hat H_{el.} |\Phi_{\bf m}\rangle.
 \ee
Hence,  the dynamics of the atomic coordinates and that of the quantum excitations  are coupled. 

The Hamiltonian $\hat H_M$ provides the time evolution of the atomic coordinates in the absence of quantum excitations and of the coupling to the heat-bath and reads  
 \be
 \hat H_M= \frac{\hat P_Q}{2M} + \hat V(Q),
 \ee
  where $V(Q)$ is a potential energy corresponding to lowest Born-Oppenheimer energy surface, $M$ is the atomic mass (here taken to be the same for all atoms, for sake of notational simplicity). 
  
 $H_{CL}$ in Eq. (\ref{Htot}) is the Caldeira-Leggett Hamiltonian (\cite{leggett-caldeira}, which  couples 
the atomic coordinates $Q=(q_1, \ldots, q_{3N})$ to a heat-bath of harmonic oscillators  
\be \label{HCL}
&& \hspace{-5mm}\hat H_{CL} = \sum_{\alpha=1}^{3 N}\sum_{j=1}^{\infty}\left(\frac{\hat \pi_j^2}{2\mu_j}+\frac{1}{2}\mu_j \omega_j^2 \hat x_j^2 - c_j \hat x_j \hat q_\alpha + \frac{c_j^2}{2\mu_j \omega_j^2}\hat q_\alpha^2\right).\nn\\  
\ee
$X=(x_1, x_2, \ldots)$ and $\Pi = (\pi_1, \pi_2, \ldots)$ are the harmonic oscillator coordinates and momenta, 
 $\mu_j$ and $\omega_j$ denote their masses and 
 frequencies and $c_j$  are the couplings between atomic and heat bath variables.   In particular, we consider the so-called  Ohmic-bath limit for the spectrum of frequencies of  harmonic oscillations (see Ref.~\cite{PIreview}).   The last term in Eq.~(\ref{HCL}) 
 is a  standard counter-term introduced to compensate the renormalization of the molecular potential energy which occurs when the heat bath variables
  are traced out.  

\section{Microscopic Quantum Field Theory for Dissipative Transport}
\label{RQFT}

The model defined in the previous section was used in Ref. \cite{Elia1} as the starting point to derive a quantum field theory which describes  the dissipative real-time dynamics of the quantum excitations, after all the molecular and heat-bath degrees of freedom in this system have been rigorously traced out from the density matrix. 

Let us consider the probability to observe at some time $t$  at the site ${\bf k}_f$ a quantum excitation, which  was initially created at $t=0$ at the site $\kk_i$:
\be
P_t({\bf k}_f,t|{\bf k}_i) = \frac{\textrm{Tr}[|{\bf k}_f\rangle \langle {\bf k}_f| \hat\rho(t)]}{\mathcal{N}}
\ee
where 
\be
\label{pcond}
\mathcal{N}= \textrm{Tr}[\hat\rho(t)]=\textrm{Tr}[e^{i H t} \hat\rho(0) e^{-i H t}],
\ee
is the normalization of the density matrix and 
\be
\hat \rho(0) = |\kk_i \rangle \langle \kk_i | \times e^{-\beta\hat H_{CL}}\times e^{-\beta \hat H_{M}}
\ee
specifies our choice for the initial condition.  
 
The Feynman-Vernon path integral representation of the probability density $P_t({\bf k}_f,t|{\bf k}_i) $ is derived by applying the Trotter formula to the two evolution operators appearing in Eq.~(\ref{pcond}). In deriving the path integral, we choose to represent  the system's state using the following basis set 
\be
| Q, X, \phi_{\bf l}\rangle = |Q\rangle \otimes |\phi_{\bf l}\rangle \otimes |X\rangle, 
\ee
where $|\phi_{\bf l}\rangle$ is the bosonic coherent state associated to  the quantum excitation, while $|Q\rangle$ and  $|X\rangle $ denotes the position eigenstates of the molecular atoms and the  set of  Caldeira-Leggett variables, respectively.
  
The functional integrals over the heat-bath Caldeira-Leggett auxiliary variables and that over small thermal oscillations of the molecule around its minimum-energy configuration can be put in a  Gaussian form, hence can be evaluated  analytically. This procedure generates an influence functional containing new effective couplings between the coherent fields associated to the quantum excitation density. This effective functional accounts for the dissipative character of the quantum transport dynamics. 

As usual, the path integral representation of the time-dependent density matrix involves the doubling of the degrees of freedom, corresponding to splitting  the Trotter representation of forward- and backward- time-evolution operators,  $e^{-i t \hat H}$ and $e^{i t \hat H}$ in Eq. (\ref{pcond}). In particular, our approach involves forward- and backward- propagating complex coherent fields, hereby denoted with $\phi^{'}_{\bf m}(t)$ and $\phi^{''}_{\bf m}(t)$, respectively. 

 As a merely formal trick to avoid using the Keldysh contour formalism, in Ref. \cite{Elia1} we proposed to interpret the fields evolving backwards in time as describing  (non-physical) ``anti-particle" excitations propagating forward in time. For notational convenience, we introduced field doublets which collectively describe both the quantum excitation and its fictitious ``anti-particle" component: 
\be
\Phi_{\bf m}(t) = (\phi'_{\bf m}(t), \phi''_{\bf m}(t)), \,\quad \bar \Phi_{\bf m}(t) \equiv \Phi_{\bf m}^\dagger (t)\tau_3,
\ee
where $\tau_3\equiv \text{diag}(1,-1)$ is the third Pauli matrix. 

The path integral  representation of the probability density $P_t({\bf k}_f,t|{\bf k}_i)$ is given by
\be\label{Pnn}
P_t({\bf k}_f,t|{\bf k}_i) &=& \frac{-1}{\mathcal{N}} \int \mathcal{D}\bar{\Phi}~ \D \Phi~ e^{-\mathcal{L}_{0}}~n_{\kk_f}(t)~n_{\kk_i}(0)\nn\\
&& \qquad ~ \cdot ~e^{\frac{i}{\hbar}S_0[\bar\Phi, \Phi]}~e^{\frac{i}{\hbar}S_{int}[\bar\Phi, \Phi]},
\ee
where 
\be
n_{\kk}(\tau) \equiv \frac{1}{2} \bar{\Phi}_{\kk}(\tau)~\tau_3 \tau_1~\Phi_{\kk}(\tau),
\ee
and $\tau_1 \equiv \left(\begin{array}{cc}0 &1\\ 1 & 0
\end{array}\right)$ is the first Pauli matrix. 
$\mathcal{L}_0$ is a surface term which arises from the over completeness of the  coherent states and reads
\be
\mathcal{L}_0 =\bar{\Phi}_{\m}(0)\tau_3\tau_+ \Phi_{\m}(0), \quad \text{with} \quad \tau_{\pm}= \frac{1}{2}\left( 1\pm \tau_3\right).
\nn\\
\ee
The terms $n_{\kk_i}(0)$ and  $n_{\kk_f}(t)$ in the path integral (\ref{Pnn}) arise from specifying the initial and final position of the quantum excitation, respectively. In the equivalent relativistic-like quantum field theory formalism, these terms contain both $\Phi$ and $\bar \Phi$ fields, hence correspond to density operators in which a particle (quantum excitation) is created and its (fictious)  anti-particle is annihilated. 
The normalization of the probability density  is given by 
\be\label{corr}
&&\mathcal{N} \equiv -\int \mathcal{D}\bar{\Phi}~ \D \Phi~ e^{-\mathcal{L}_{0}}~n_{\kk_i}(0)~e^{\frac{i}{\hbar}S_0[\bar\Phi, \Phi]}~e^{\frac{i}{\hbar}S_{int}[\bar\Phi, \Phi]}.\nn\\
\ee

The action functionals appearing at the exponent of Eq.~(\ref{corr}) are defined as follows:
 \be
\label{def2}
&&S_{0}[\bar \Phi, \Phi] = \int_0^t dt'~\bar{\Phi}_{\bf m} ~( i \hbar \partial_{t'} ~\delta_{\bf m n}  - f^0_{\bf m n})~\Phi_{\bf n},\\
\label{def3}
&&S_{int}[\bar \Phi, \Phi] = \frac{1}{4} \int_0^t d t'dt''~ \bar{\Phi}_{\bf l}(t') ~\tau_3 f^a_{\bf l h}~\Phi_{\bf h}(t') \nonumber\\
&& \mathcal{V}_{a b}(t'-t'')~\bar \Phi_{\bf m}(t'')f^b_{\bf m n}~\Phi_{\bf n}(t'') 
+ \frac {i M \gamma }{ \beta \hbar} \int_0^t dt' dt'' \nn\\
&& \bar{\Phi}_{\bf l}(t') ~f^a_{\bf l h}\Phi_{\bf h}(t')  \Delta_{a b}(t'-t'') \bar{\Phi}_{\bf m}(t'') ~f^b_{\bf m n}~\Phi_{\bf n}(t''),\nn\\
\ee
where $\beta=1/k_BT$. The indexes $a,b$ run over the $3  N$ components of the molecular configuration vector $Q$.  

The explicit expression of the Green's functions $ \Delta_{ab}(t-t')$ and  $\mathcal{V}_{ab}(t-t')$, which mediate the interaction of the hole with  the molecular vibrations and with the viscous heat-bath are given in Ref.~\cite{Elia1} and depend on the viscosity $\gamma$ and molecular normal modes and frequencies.  

The matrix elements $f_{\m \n}^0$ and $f^a_{\m \n}$ are defined as 
\be \label{hopp_param}
f_{\m \n}^0 = f_{\m \n}[Q_0], \qquad f^a_{\m \n} = \frac{\partial}{\partial Q^a} f_{\m \n}[Q_0],
\ee
 where $Q_0$ is the equilibrium molecular configuration in the absence of quantum excitations. 

In Ref. \cite{Elia1} this QFT  was used to analyze the effects of dissipation on the quantum transport, using perturbation theory.  Indeed, in the short-time regime,  it is possible to develop Feynman diagram techniques which immediately yield the corrections to the unitary (i.e. free) quantum propagation of the excitation. The leading-perturbative correction  corresponds to simple analytical formulas. This approach was also used  to study the loss of quantum coherence induced by the coupling of the excitation to the environment. 

This relativistic-like QFT description can in principle be used to investigate also the dynamics over long time intervals.  However, in this case,  the correlation functions are dominated by multiple scattering with the damped molecular oscillations, hence  must be evaluated non-perturbatively, for example by means of a saddle-point approximation. While developing such a scheme is in principle possible, it is leads to microscopic self-consistent equations which depend on a large number of effective parameters and are computationally  challenging to solve. In view of this difficulties, in the next sections we shall develop a much simpler EFT which describes the same physics at low space-time resolution power, i.e. in the long-distance and long-time limit. As we shall see, such a theory turns out to be much simpler and depends only on a few parameters and even admits analytic solutions.  

\section{EFT for Dissipative Quantum Transport}
\label{long-time}
We are interested in  constructing an EFT which describes the same IR dynamics of the microscopic theory defined in the previous section, at a much lower level of spatial and temporal resolution. In particular, we focus on macromolecular systems for which the quantum excitation can cover distances which are long compared to those at which the molecular three-dimensional structure is resolved. In Fourier space, this implies that the corresponding coherent fields have only soft momentum components. Consequently, the effective interaction terms with a larger and larger number of spatial derivatives are increasingly irrelevant.  Similarly, since thermal oscillations are damped by the coupling with the heat-bath, in order to investigate the  long-time regime, the terms with higher number of time derivatives can be dropped. 

Let us now define the action functional of the effective theory.  We begin by analyzing the kinetic term and we first consider the case in which the hopping of the excitations between molecular orbitals which are spatially neighboring. In this case, the hopping matrix in the microscopic theory defined in Eq.~(\ref{def2}) takes the simple form 
\be
f_{\n \m}(Q ) &=&\sum_{\hat{i}}  \tau_{\n \m}(Q) (\delta_{\m (\n+\hat{i})} + \delta_{\m (\n-\hat{i})}) \nn\\
&-&  e^0_{\n}(Q) ~\delta_{\n \m}.
\ee
where the sum runs over the unit vectors $\hat{i}$ pointing to the nearest neighbor sites from the site $\m$.  

Due to the low spatial resolution, in the corresponding effective theory we can replace the discrete site indexes $\m,\n$ with continuous variables $\x,\y$, i.e.  to introduce continuous complex field doublets $ \Phi_{\bf n}(t) \to \Phi(\x,t), \quad  \bar\Phi_{\bf n}(t) \to \bar\Phi(\x,t)$. 
In the continuum limit, the matrices $f^0_{\m \n}$ and $f^{i}_{\n \m}$ become differential operators
\be
\label{f01}
f^0_{\m \n} &\to& \delta(\x-\y)\left[ \epsilon(\x) - \frac{\hbar^2~\mu^{-1}_{i j}(\x)}{2}  \partial_i \partial_j\right], \\
\label{f02}
f^{a}_{\n \m} &\to&\delta(\x-\y)\left[\epsilon^a(\x) - \frac{\hbar^2\mu^{a\, -1}_{i j}(\x)}{2}   \partial_i \partial_j\right].
\ee
Notice that $\mu_{i j}({\bf x}) $ can be interpreted as a position dependent effective mass tensor. 

With the notation defined above, the free component of the action functional $S_0$ is written as
\be\label{S0}
S_0[\bar \Phi, \Phi] &\simeq& \int_0^t dt' \int d \x ~\bar \Phi(\x, t')~ \left( i\hbar \partial_t  - \epsilon(\x) \right.\nn\\
&+&\left. \frac{\hbar^2 }{2} \mu^{-1}_{ij}(\x) \partial_i  \partial_j\right)~\Phi(\x,t').
\ee
It is immediate to verify that accounting for non-nearest neighbor hopping leads to higher derivative terms. According to our power-counting scheme, 
such terms are irrelevant in the IR limit and can be ignored. 

Let us now analyze the interaction terms in Eq.~(\ref{def3}). The Green's functions $ \Delta_{ab}(t-t')$ and  $\mathcal{V}_{ab}(t-t')$ are evaluated explicity in Ref. \cite{Elia1}, where they are shown to decay exponentially at time-scales of the order of the inverse collision rate $t\sim 1/\gamma$. Since we are focusing in time intervals $t\gg 1/\gamma$, they can be replaced by 
\be\label{expa}
 \Delta_{ab}(t-t') &\simeq& d^{(0)}_{ab} \delta(t-t') + d^{(1)}_{ab} ~i \hbar~\frac{d}{dt} \delta(t-t')+ \ldots,\nn\\
 \mathcal{V}_{ab}(t-t') &\simeq& v^{(0)}_{ab} \delta(t-t') + v^{(1)}_{ab} ~i \hbar \frac{d}{dt} \delta(t-t')+ \ldots \nn \\
\ee
Corrections to these terms are irrelevant, as involve higher time-derivatives. 
The effective interaction becomes
\be\label{Sint}
&& \hspace{-3mm} S_{int}[\bar\Phi,  \Phi] \simeq \frac{1}{4}\int dt'\int dt''\int d\x\left[\bar\Phi(\x,t')\tau_3\left(\epsilon^a(\x) - \right.\right.\nn\\
&& \hspace{-3mm} \left. \left.\frac{\hbar^2 }{2 } \mu^{a~-1}_{i j}(\x) \partial_i \partial_j\right) \Phi(\x,t')\left[\left(v^{(0)}_{a b}- i v^{(1)}_{ab} \frac{d}{dt'} \right) \delta(t'-t'')\right]
\right.\nn\\
&& \hspace{-3mm} \left. \bar\Phi(\x,t'')  \left(\epsilon^b(\x) - \frac{\hbar^2 }{2} \mu^{b~-1}_{i' j'}(x)  \partial_{i'} \partial_{j'}\right) \Phi(\x,t'')\right] + \frac{iM\gamma}{\beta \hbar} \nn \\  
&& \hspace{-3mm} \int dt' \int dt'' \int d\x\left[~\bar\Phi(\x,t')\left(\epsilon^a(\x) - \frac{\hbar^2 }{2} \mu^{a~-1}_{i j}(\x)  \partial_i \partial_j \right) \right.\nn \\
&& \hspace{-3mm} \left. \Phi(\x,t') \left[\left(d^{(0)}_{a b} - i d^{(1)}_{ab} \frac{d}{dt'} \right)~\delta(t'-t'')\right]\bar\Phi(\x,t'') \left(\epsilon^b(\x) \right.\right. \nn\\
&& \hspace{-3mm} \left. \left. - \frac{\hbar^2 }{2 } \mu^{b~-1}_{i' j'}(\x)  \partial_{i'} \partial_{j'}\right)\Phi(\x,t'')\right] .\nn\\
 \ee
The combination $D = 1/(\beta M \gamma)$, yields the diffusion coefficient of the atoms in their surrounding heat-bath. 
 
The scalar and tensor fields $\epsilon(\x), \mu_{ij}(\x)$ appearing in Eq.~(\ref{S0})  encode the information about the  conformational and electronic structure of the molecule in the neighborhood of the point $\x$, while the uniform tensors $v^{(k)}_{ab}$ and $d^{(k)}_{ab}$ (with $k=0,1$) parametrize the fluctuation-dissipation effects arising from the coupling of the molecule with its heat-bath. It is important to emphasize that, even though these quantities do not depend explicitly on time, they are the result of pre-averaging out the small thermal oscillations, hence encode dynamical effects.  


The effective action (\ref{Sint}) does not yet define our EFT. Indeed, so far, we have only performed a continuous formulation of the original microscopic theory, and taken the so-called Ohmic bath limit of the Caldeira-Leggett model.  On the other hand, we have now yet exploited the  decoupling of UV and IR length-scales and taken the large-distance limit. 

To derive the effective couplings in the EFT, let us consider the Fourier transform of the space-dependent scalar and tensor parameters  in Eq.~(\ref{Sint}) (in the following we focus on $\epsilon(\x)$ for sake of definiteness):
\be
\epsilon ({\bf p}) = \epsilon_0 ~\delta({\bf p}) + \delta \epsilon({\bf p}). 
\ee
We note that $\epsilon_0$ is the Fourier component which corresponds to a uniform field.
We recall that the parameter fields are assumed to vary over length scales which are much shorter than those over which the quantum excitation's density changes. In Fourier space, this implies that $\epsilon({\bf p})$ has only hard components
 \be
 \delta\epsilon({\bf p})\simeq 0, \qquad \text{for} \quad|{\bf p}|\lesssim 1/\lambda,\ee
 while the field $\Phi({\bf p})$ has only soft Fourier components,
 \be
 \Phi({\bf p})\simeq 0, \qquad \text{for} \quad |{\bf p}|\sim 1/\lambda.
 \ee
 Due to such a  decoupling,  all the short-distance local fluctuations of the parameter fields average away:
\be
\int d\x \epsilon(\x) \bar \Phi(\x, t) \Phi(\x, t) &\simeq&  \epsilon_0 ~\int d\x\bar \Phi(\x, t) \Phi(\x, t)  \nn\\
 \ee
since 
\be
&&\int d\x \delta \epsilon(\x) \bar \Phi(\x, t) \Phi(\x, t)  =
\nn\\
&=& \int \frac{d{\bf p}}{2\pi} \int \frac{d{\bf q}}{2\pi} \bar \Phi({\bf p}) \delta \epsilon({\bf q}) \Phi(-({\bf p}+{\bf q}),t)  \simeq 0. 
\ee
In addition, all coupling terms in Eq.~(\ref{Sint}) which include spatial derivatives of the fields are irrelevant in the large distance limit, hence can be neglected. Similarly, for the effective mass tensor field, one has 
\be
\mu_{i j}({\bf x}) \simeq m_{ij}.
\ee 
 
In conclusion, to the lowest-order in our power-counting scheme,  the path integral which represents the excitation probability density reduces to
\be\label{EFtheory}
P({\bf y}, t|{\bf x})  &=& \frac{-1}{\mathcal{N}}\int \D \bar \Phi \D \Phi~n(\y, t)~n(\x, 0)\nn\\
&&\cdot~ e^{-\mathcal{L}_1
+\frac{i}{\hbar} S^{eff}_0[\bar \Phi, \Phi]+ \frac{i}{\hbar} S_{int}^{eff}[\bar \Phi, \Phi]},\nn\\
\ee
where 
\be
\mathcal{N}  &\simeq& \int \D \bar \Phi \D \Phi~n(\x, 0)~
 e^{-\mathcal{L}_1
+\frac{i}{\hbar} S^{eff}_0[\bar \Phi, \Phi]}\nn\\
&&\cdot~ e^{\frac{i}{\hbar} S_{int}^{eff}[\bar \Phi, \Phi]},\nn\\
\mathcal{L}_1[\bar \Phi, \Phi]  &=& \int d{\bf x} ~\bar{\Phi}(\x, 0)\tau_3\tau_+ \Phi(\x,0),\\
S^{eff}_0[\bar \Phi, \Phi] &=& \int_0^t dt' \hskip-1.5mm\int \hskip-0.5mm d\x ~\bar \Phi\big{[}i \hbar \partial_{t'} -  \epsilon_0 +\frac{\hbar^2}{2 } m_{ij}^{-1} \partial_i \partial_j \big{]}\Phi, \nn\\\\
\label{SintEFT}
S_{int}^{eff}[\bar \Phi, \Phi]&=&\int_0^tdt'\int d\x ~\Big{[} \bar \Phi \Phi \left(A_v^0 -  \hbar A_v^1i \partial_{t'}\right) \bar \Phi \tau_3 \Phi\nn\\
 && + \frac{i}{D \hbar \beta^2}\bar \Phi  \Phi\left(A_d^0 - \hbar A^1_d i \partial_{t'}\right) \bar \Phi \Phi \Big{]}.
\ee
$A_v^0,A_v^1, A_d^0$ and $A_d^1$ are real effective coupling constants. 
It is important to emphasize that the couplings of  low-energy effective theories may depend in general on the heat-bath temperature~\cite{LeBellac}.  

It is important to emphasize that the effective action defined so far only describes the long-time and large-distance dynamics of the excitations. Indeed, the short-distance and short-time physics has been quenched, by introducing the cut-offs  on the length and time scales and by neglecting the high derivative terms in the effective action. 

In general, completely neglecting the high-frequency modes represents a very crude approximation, since  the short-distance physics does have an influence on the long-distance dynamics. The crucial point to make is that, at a low-resolution power, the long-distance dynamics becomes insensitive to the \emph{details} of the short-distance physics, which can be therefore encoded by  by means of local effective interactions.  

These physical ideas are implemented by the renormalization procedure. In practice, one adds to the effective action in Eq.(\ref{EFtheory}) the \emph{most general} set of \emph{local} effective vertexes compatible with the symmetry of the underlying microscopic theory and in retains only the terms among them which display the least number of derivatives and fields (see e.g. the discussion in \cite{EFT1}).  
 
However, in our specific case, the renormalization procedure is in fact redundant. Indeed, we observe that the effective Lagrangian in Eq.~(\ref{EFtheory}) already contains all  possible local effective couplings with the least number of derivatives and fields.  Hence, to lowest-order, the renormalization procedure simply amounts to rescaling the corresponding effective parameters $A_v^0,A_v^1, A_d^0$ and $A_d^1$. 
\newline

Let us summarize what we have obtained so far. We have shown that, in the large-distance and long-time limit, the probability density for the quantum excitations exciton can be \emph{formally} mapped into a vacuum-to-vacuum two-point function in a relativistic-like quantum field theory with local 4-field interactions. Note that our  theory contains an imaginary coupling constant, which breaks unitarity and describes the dissipation generated by the coupling with the damped molecular oscillations. 




\section{Effective Stochastic Description}
\label{effstoch}
In this section we analyze the dynamics of the quantum excitation in the asymptotic long-time regime. In this limit, the large number of collisions with the molecular vibrations and with the environment depletes the quantum coherence. As a result, the emergent dynamics of the quantum excitation is diffusive and quasi-classical.
 
In order to investigate the quasi-classical limit it is convenient to switch to the first-quantization formalism and represent the time-dependent  probability (\ref{EFtheory}) using the coordinate representation path integral.

To this end, it is important to recall that the position eigenstates of the EFT do not coincide with those of a fundamental theory. Indeed,   
in the EFT, each  point {\bf X} is indistinguishable from those which lie in a neighborhood of the size of the probe's resolution power, $\lambda$. 
Most generally, the position eigenstates in the EFT,  $|{\bf R}\rangle_\lambda$  are defined as
\be \label{regularize}
|{\bf R}\rangle_\lambda \equiv \int d{\bf X} ~\Psi_{\lambda}({\bf X}-{\bf R})~|{\bf X}\rangle
\ee
where $|{\bf X}\rangle$ denote the position eigenstates of the microscopic theory. 
The wave function $\Psi_{\lambda}({\bf X}-{\bf R})$ is determined by the normalization condition,
\be
{}_{\lambda}\langle {\bf R}' | {\bf R}\rangle_\lambda &=&  \int d{\bf X} \Psi_{\lambda}({\bf X}-{\bf R})
~\Psi^{*}_{\lambda}({\bf X}-{\bf R}')\nn\\
&=& \delta_\lambda({\bf R}'-{\bf R}) 
\ee
where $\delta_{\lambda}({\bf R'}-{\bf R})$ is some smeared representation of the $\delta$-function, of width $\lambda$.
In particular, in the following, we adopt a Gaussian smearing (in three dimensions) 
\be\label{smearing}
\delta_{\lambda}({\bf R}-{\bf R}')& \equiv& \sqrt{\det m}\left(\frac{1}{2 \pi~\text{Tr}\,m~ \lambda^2}\right)^{3/2} \nn\\
&& \times ~e^{-\frac{(R_i-R'_i) m_{ij}(R_j-R'_j) }{2 \text{Tr}~m~\lambda^2}}.
\ee
With this choice, the effective wave-function $\Psi_\lambda({\bf X})$ reads:
\be
\Psi_\lambda({\bf X}) = \sqrt{\det m} \left(\frac{1}{\pi~ \text{Tr} m~ \lambda^2}\right)^{3/2} ~e^{-\frac{ X_i m_{ij} X_j}{ \text{Tr}\,m~\lambda^2}}.
\ee
Notice that this regularization choice corresponds to approximating the local delocalization with a harmonic oscillator wave-function and takes into account for the tensor structure of the effective mass. However, the specific  short-distance structure of this wave function is irrelevant for the long-distance dynamics. 

Denoting with ${\bf X}[\tau]$ and ${\bf Y}[\tau]$ the paths in coordinate space of  quantum excitation described by the coherent 
fields $\phi^{'}$ and $\phi^{''}$, respectively,  the probability density can be written as: 
\be
\label{Pfirst}
P({\bf y}, t| {\bf x},0) &=& \int \D {\bf X} \int \D {\bf Y} ~e^{\frac{i}{\hbar} \int_0^t d\tau \frac{1}{2}\,m_{ij}\,(\dot X_i \dot X_j-\dot Y_i \dot Y_j)} \nn\\
&&~\times e^{\frac{i}{\hbar} \left(I[{\bf X},{\bf Y}]+ J[{\bf X}, {\bf Y}]\right)}, 
\ee 
The functionals $I[{\bf X},{\bf Y}]$ and $J[{\bf X},{\bf Y}]$ originate from   translating into the first quantized representation the  field-theoretic functional $S_{int}$ appearing in Eq. (\ref{SintEFT}). In appendix \ref{translation}, we derive them explicitly and obtain $J=0$ and $I=I_1+I_2+I_3$, with
\be
&&\hspace{-0.5cm}I_1 = I_2 =  \frac{i t}{\beta^2 D \hbar} \frac{\sqrt{\det m}}{ (4 \text{Tr}\,m~\lambda^2 \pi)^{3/2}}~A_d^0\\
&&\hspace{-0.5cm} I_3 = \frac{\sqrt{\det{m}} }{\beta^2 D \hbar\, ( 4\,\text{Tr}\,m~\pi \lambda^2)^{3/2}} \int_0^t dT\left\{ e^{\frac{-m_{ij}}{4 \text{Tr} m\, \lambda^2}(X- Y)_i (X- Y)_j}\right. \nn\\
&& \left. \left(-i 2 A_d^0  + \frac{ \hbar A_d^1}{2 \text{Tr}\,m~\lambda^2}~ m_{ij}~( Y- X)_i (\dot Y+ \dot X)_j\right) \right\}.
\ee

We now perform the following change of variables in the path integral:
\be
{\bf R} &=& \frac{1}{2} ({\bf X+Y})\\
{\bf Q} &=& {\bf X-Y}.
\ee
In addition, for reasons which will become clear soon, it is convenient to introduce  the following tensor combination:
\be\label{DandGamma}
\Gamma^{0}_{ij} &=& \frac{A_d^1 \sqrt{\det m}}{16 D \beta^2 (\text{Tr}\, m )^{5/2}\,\lambda^5 \sqrt{\pi^{3} }}~m_{i j}.
\ee

Using (\ref{DandGamma}), the path integral in Eq.~(\ref{Pfirst}) reads
\be\label{coordPI}
&&P({\bf y}, t| {\bf x},0) =~e^{-4 t \frac{\lambda^2  \text{Tr}~\Gamma^0 A_d^0}{\hbar^2\beta A_d^1}} \int_{\bf x}^{\bf y} \mathcal{D}~{\bf R} \int_{\bf 0}^{\bf 0}\mathcal{D}{\bf Q}\nn\\
&& e^{ \int_0^t dT\left\{ \frac{i}{\hbar} ~m_{ij}  \dot R_i \dot Q_j + \left(\frac{4\,\lambda^2 \text{Tr} \Gamma_0 A_d^0}{\hbar^2  \beta A_d^1} -\frac{ i \Gamma^0_{ij}}{\hbar}Q_i \dot R_j\right) e^{-\frac{m_{ij} Q_j Q_i }{4 \text{Tr}m~ \lambda^2}}
 \right\} } 
  \nn\\
\ee
We now observe that, by expanding the exponent in Eq. (\ref{coordPI}) to $\mathcal{O}({\bf Q^2})$ we obtain:
\be\label{coordPI3}
&&P({\bf y}, t| {\bf x},0) = \int_{\bf x}^{\bf y} \mathcal{D}~{\bf R} \int_{\bf 0}^{\bf 0}\mathcal{D}{\bf Q}\nn\\
&& e^{ \int_0^t dT\left\{\frac{i}{\hbar} ~m_{ij}  \dot R_i \dot Q_j -\frac{\Gamma_{i j} A_d^0}{\hbar^2  \beta A_d^1} Q_j Q_i  - \frac{ i \Gamma^0_{ij}}{\hbar}Q_i  \dot R_j
\right\} } \nn\\
\ee
Performing the  Gaussian functional integration over ${\bf Q}$ one obtains (neglecting as usual any overall multiplication constant)
\be\label{diffusivePI}
&&P_{cl.}({\bf y}, t| {\bf x},0)=\int_{\bf x}^{\bf y}\mathcal{D}{\bf R}\nn\\
&& e^{-\frac{\beta^2}{4}
\int_0^t dT \left(m_{i l} \ddot  R_l + \Gamma_{i l}^{0} \dot R_l  \right) D_{ij}^{0}~\left(m_{jk}\ddot  R_k + \Gamma_{j k}^{0} \dot R_k  \right)}.
  \nn\\
\ee
where we have introduced the tensor
\be
D^0_{ij} \equiv \frac{1}{\beta}~\left(\frac{A_d^1}{\beta A^0_d}\right)~\Gamma_{ij}^{0\, -1}.
\ee

We  now recognize that Eq. (\ref{diffusivePI}) has the same structure of the  Onsager-Machlup functional integral representation~\cite{OMfunctional} of the solution of a Fokker-Planck equation with an anisotropic viscosity tensor $\Gamma_{i j}^0$ and diffusion tensor $D^{0}_{ij}$. The request that the system should ultimately reach a thermal equilibrium with the surrounding heat-bath implies the fluctuation-dissipation relationship,
\be\label{fluct-diss}
D_{ij}^{0} = \frac{1}{\beta} \Gamma_{ij}^{0 -1}
\ee
This condition determines a relationship between the effective parameters, $A_d^1 =  \beta~A_d^0.$

The  friction tensor $\Gamma_{i j}^0$ has to be determined by  matching of the predictions of our EFT against the quantum excitation's, either measured  experimentally ~\cite{measuring-mobility} or computed theoretically using a (more) microscopic model \cite{Troisi}. This renormalization procedure will be illustrated in detail in section \ref{solutionandrenormalization}.

Let us now return to the full path integral (\ref{coordPI}) in order to determine the quantum corrections to Eq. (\ref{diffusivePI}). For sake simplicity, in the rest of this work we focus in the high-friction limit, in which the dynamics is over-damped. In addition, without loss of generality, we can assume that the friction tensor $\Gamma^0_{ij}$ is diagonal. 
and introduce the inner product notation
\be\label{inner}
{\bf A}\cdot {\bf B}  \equiv g_{ij}^0~ A_i B_j,
\ee
where $g_{ij}^0$ is a diagonal metric tensor defined as
\be
\Gamma_{ij}^0 = \frac{1}{\beta D_0}~g^0_{i j},
\ee
and $D_0$ has the dimension of a diffusion constant. 

Dropping the inertial term, the path integral (\ref{coordPI}) reduces to
\be
\label{coordPI2}
&&\hspace{-0.5 cm} P({\bf y}, t| {\bf x},0) = \int_{\bf x}^{\bf y} \mathcal{D}~{\bf R} \int_{\bf 0}^{\bf 0}\mathcal{D}{\bf Q}~e^{ -\int_0^t dT\left\{ \frac{{\bf Q}^2}{\hbar^2 D_{0} \beta^2} + \frac{ i {\bf Q}\cdot \dot{\bf R}}{\hbar \beta D_{0} } \right\} }\nn\\
&& e^{ \int_0^t dT\left\{ \left(\frac{4\lambda^2}{\hbar^2 D_{0} \beta^2} - \frac{ i {\bf Q}\cdot \dot{\bf R}}{\hbar \beta D_{0}}\right) V({\bf Q})
+ \frac{ i {\bf Q}\cdot \dot{\bf R}}{\hbar \beta D_{0}}\frac{\bf Q^2}{4 \lambda^2} \right\}} 
  \nn\\
\ee
where 
\be\label{VQ}
V({\bf Q})= e^{-\frac{\bf Q^2}{ 4 \lambda^2}}-1+\frac{\bf Q^2}{ 4 \lambda^2}.
\ee
The probability distribution (\ref{coordPI2}) can be cast in the following convenient form,
\be
\label{QL}
&&P({\bf y}, t| {\bf x},0) = \int_{\bf x}^{\bf y} \mathcal{D}{\bf R}\, e^{-S_{eff}[\dot {\bf R}]}
\ee
where the effective ``action" functional $S_{eff}$ is defined as
\be\label{Seff}
&&\hspace{-0.5cm} e^{-S_{eff}[\dot {\bf R}]} \equiv \int_{\bf 0}^{\bf 0}\mathcal{D}{\bf Q}~e^{ -\int_0^t dT\left\{ \frac{{\bf Q}^2}{\hbar^2 D_0 \beta^2} 
+ \frac{ i {\bf Q}\cdot \dot{\bf R}}{\hbar \beta D_0 } \right\}} \nn\\
&&\times e^{ \int_0^t dT \left(\frac{4\lambda^2}{\hbar^2 D_0 \beta^2} - \frac{ i {\bf Q}\cdot \dot{\bf R}}{\hbar \beta D_0}\right) V({\bf Q}) +\frac{ i {\bf Q}\cdot \dot{\bf R}}{\hbar \beta D_0}\frac{\bf Q^2}{4 \lambda^2}}.\nn\\
\ee
 We emphasize that, at this level, no approximation has been made on the path integral (\ref{EFtheory}). 
Hence, Eq.~(\ref{QL}) represents the full real-time dynamics of the quantum excitation in our  EFT, to leading-order in the momentum-frequency power counting scheme.

\section{Computing the Quantum Effective Functional}
\label{QEFTfunc}

In general, the effective action  $S_{eff}[\dot{\bf R}]$ which enters Eq. (\ref{Seff}) is a non-local functional of the path ${\bf R}(\tau)$ and such a non-locality reflects the quantum delocalization of the excitation's wave-function. However, we shall now show that, in the limit of low-spatial resolution, the effective action $S_{eff}$ can be systematically represented as an expansion in local functionals. 
In this regime, the quantum dynamics of our EFT can be described as a modified diffusion process. 

To derive this result, we begin by recalling that in a thermal heat-bath the amplitude of quantum fluctuations are of the order of the De Broglie's thermal wave-length 
\be\label{lambdaB}
\lambda_B \equiv \hbar \sqrt{\frac{\beta}{2 \pi \mu_0}},\ee
where $\mu_0= \frac{1}{3}\text{Tr} ~m_{i j}$ is an effective mass scale for the quantum excitation. 

 If the cut-off scale $\lambda$ is chosen in such  a way that $\lambda \gg \lambda_B$,  one has
\be\label{power-counting}
\xi = \frac{\lambda_B}{\lambda}   \ll 1\quad \text{and} \quad \frac{|{\bf Q}|}{\lambda} \sim \xi.
\ee

Hence, $\xi$ provides a small expansion parameter which enables us to evaluate the effective action $S_{eff}$ within a systematic perturbation theory. We note that, in order to obtain the   $\mathcal{O}(\xi^2)$ expression,  it is sufficient to expand the exponent in the second line of Eq. (\ref{Seff}) to  order in ${\bf Q}^4$, since all higher-order terms lead to corrections which are of order  $\mathcal{O}(\xi^4)$. 
After discretizing the time interval $t$ in $N_t$ steps, the path integral factorizes as a product of $N_t$ moments of Gaussian distribution, in the form:
\be\label{Gaussian}
&&\hspace{-5mm} \int d{\bf Q}_k  \exp\left[i\frac{ ({\bf R}_{k+1}-{\bf R}_{k})}{D_0 \sqrt{2 \pi \beta \mu_0}} \cdot \frac{{\bf Q}_k}{\lambda_B } -
\frac{\Delta t}{2 \beta D_0\mu_0 \pi} \ \frac{{\bf Q}^2_k}{\lambda_B^2 }\right] \nn\\
&&\hspace{-5.5mm} \left(1 - \frac{i ({\bf R}_{k+1}-{\bf R}_{k}) \cdot {\bf Q}_k }{4 \beta D_0 \sqrt{ 2\pi \mu_0 \beta}\, \lambda_B} \ 
\frac{{\bf Q}_k^2}{\lambda^2} + \frac{\Delta t ~ {\bf Q}^2}{16 \beta D_0 \mu_0 \pi \, \lambda_B^2} \ \frac{{\bf Q}^2}{\lambda^2} \right),\nn\\\ee 
where  ${\bf Q}_k \equiv {\bf Q}(t_k)$.

The incremental time interval $\Delta t\equiv t/N_t$ plays the role of a regularization cut-off  and is the time analog of the distance regularization cut-off $\lambda$. We recall that, in the  EFT framework, both $\Delta t$ and $\lambda$  are kept  finite  at all stages of the calculation.

The result of the Gaussian integral (\ref{Gaussian}) can be written in the form
\be
&& \hspace{-4mm}  \mathcal{N} e^{-\frac{({\bf R}_{k+1}-{\bf R}_{k})^2}{4 D_0 \Delta t}}
~e^{\frac{ \beta \mu_0 \pi}{16 \Delta t^3 }\xi^2\left( 5 \Delta t ({\bf R}_{k+1}-{\bf R}_{k})^2- 3 \frac{ ({\bf R}_{k+1}-{\bf R}_{k})^4}{4 D_0} \right)}, \nn\\
\ee
where $\mathcal{N}$ is an irrelevant constant factor.
Multiplying all the $N_t$ terms and restoring the continuum notation, we obtain
\be\label{S_diff_quant}
S_{eff} \simeq \int_0^t dT \left[\frac{\dot{\bf R}^2}{4 D^b_{0}} - \xi^2 \left( C^b_2~\dot{\bf R}^2 - C^b_4~\dot{\bf R}^4\right)\right], \, \quad 
\ee
where the coefficients of the correction terms order $\mathcal{O}(\xi^2)$ are
\be\label{C2}
C^{b}_2 &\equiv& \frac{ 5  \beta \mu_0 \pi}{16\, \Delta t }, \qquad
C^{b}_4 \equiv \frac{ 3  \beta \mu_0 \pi}{64\, D^b_{0}}.
\ee
In Eq.s (\ref{S_diff_quant}) and (\ref{C2}),  we have added the superscript ``b'' to emphasize that $D_0^b$, $C^{b}_2$ and $C^{b}_4$ are \emph{bare} effective constants. 

Summarizing, up to corrections of order $\mathcal{O}\left(\frac{\lambda_B}{\lambda}\right)^4$, the exciton's probability density can be approximated as 
\be\label{modifiedDiff}
\hspace{-0.1mm} P({\bf y}, t| {\bf x},0) &\simeq&  \int_{\bf x}^{\bf y} \hspace{-0.5mm}\mathcal{D}{\bf R}~e^{-\int_0^t dT\left[ \frac{1}{ 4 D^b_{0}}\dot{\bf R}^2
-\xi^2 \left(C^b_2~ \dot{\bf R}^2 -  C^b_4~ \dot{\bf R}^4 \right) \right]} \nn\\
\ee

The path integral (\ref{modifiedDiff}) describes the propagation of the quantum excitation as a modified diffusion process and represents one of the main results of this paper. 

An important question is what corrections to the effective theory are needed in order to reach an accuracy of order $\xi^4$.
In effective field theories this question is addressed by including higher order terms  in the derivative expansion and defining a power-counting rule.  We emphasize that, in deriving our long-distance expression for the path integral, we have truncated two independent expansions: (i) we have retained  only the lowest terms in the spatial- and time- derivative expansion of all the fields in Eq.s (23) through (30) and (ii) we have kept only the leading-order terms in the expansion in $\xi^2$ of the first quantized path integral (\ref{coordPI2}). Furthermore, unlike in relativistic effective field theories, the cut-offs defining the time and spatial resolutions are not directly related.
Clearly, the existence of multiple expansions and cut-offs offers several alternatives for the power-counting schemes. A systematic analysis of all these possibilities is quite involved and goes beyond the scope of this work. 
Here, we limit ourselves to note that by simply retaining the next order in ${\bf Q^2}$ in the expansion of the function $V({\bf Q})$ in Eq. (\ref{VQ})  and then truncating the resulting effective action to order $\xi^4$ one may miss some important contributions, if the fields vary sufficiently rapidly in time and space. Indeed, additional  first-quantized operators at order in $\xi^4$ may generated by expanding the field operators in Eq.s (23) through (30) to include terms with higher number of space and time derivatives.

In section \ref{application}, we present an application of this framework to investigate hole transport in a long DNA molecule. We find that the lowest-order analytic effective theory gives results which are essentially indistinguishable from those obtained from numerical simulations in a more microscopic model. This finding suggests that, in practice, the inclusion of $\xi^4$  terms may not be crucial in order to achieve an accurate  description of the  long-distance long-time physics in realistic macromolecular systems. 

Finally, we emphasize that, even though the expansion in $\xi^2$ generates terms with increasing powers of $\hbar^2$, the EFT expansion is not conceptually equivalent to the semi-classical approximation \cite{QL1, QL2,QL3,QL4}. Indeed, the EFT approach is defined in terms of external cut-off scales, which set the resolution power of the theory and are chosen \emph{a priori}.

\section{Solution of the Path Integral and Renormalization}\label{RenEFT}
\label{solutionandrenormalization}

The effective theory defined in Eq.~(\ref{modifiedDiff}) explicitly depends on the cut-off scales $\Delta t$ and $\lambda$ and needs to be renormalized. This can be done by introducing appropriate counter-terms into the effective action and matching the prediction of the effective theory against experiment or more microscopic calculations, at some time-scale $t^*$. Through such a renormalization procedure, the power-law dependence of the effective coefficients on the cut-offs  $\Delta t$ and $\lambda$ is removed and is replaced by a much weaker dependence on the renormalization scale $t^*$.

To implement this program, let us consider for sake of simplicity the simple case of isotropic diffusion (i.e. $g^0_{ij}= \delta_{i j}$). The same procedure can be straightforwardly applied to the general case of anisotropic diffusion, by repeating the same analysis component-by-component. 




After introducing the renormalizing counter-terms, the path integral (\ref{modifiedDiff}) is modified as follows: 
\be\label{RenP}
\hspace{-0.1mm} P({\bf y}, t| {\bf x},0) &\simeq&  \int_{\bf x}^{\bf y} \hspace{-0.5mm}\mathcal{D}{\bf R}~e^{-\mathcal{S}_{eff} \left[\dot{ \bf R} \right] + \xi^2 \left(Q_2~ \dot{\bf R}^2 + Q_4~ \dot{\bf R}^4 \right)} , \nn\\
\ee
where $Q_2$ and $Q_4$ are  insofar unknown coefficients. To order $\xi^2$ the renormalized expression for the effective action then reads 
\be
\bar{ \mathcal{S}}_{eff} = \int_0^t dT \left[\frac{\dot{\bf R}^2}{4 D_{ren}} +  C_{ren} \dot{\bf R}^4  \right], \, \quad 
\ee
where $D_{ren}$, and $C_{ren}$ are the renormalized coefficients. 
In the following, we show how they can be determined up to  $\mathcal{O} \left(\xi^2\right)$ accuracy. 

To this end,  we first analytically compute  the  path integral given in Eq.~(\ref{RenP}) to leading-order in a perturbative expansion in  $\xi^2$. We obtain
\be
\label{probREN}
&&P({\bf y}, t| {\bf x},0) \simeq P_0 ({\bf y}, t| {\bf x},0; D^b_{0})~\left[1 + \xi^2 (C_2^b +Q_2)\right.\nn \\ 
&&\left.\cdot\left( \frac{({\bf x-y})^2}{t} - 6 D^{b}_{0}\right)- \xi^2 (C_4^b-Q_4) \left(\frac{({\bf x-y})^4}{t^3 } -  \right. \right. \nn\\
&&\left. \left.\frac{\Delta t-t}{\Delta t} ~\frac{20 D^{b}_0 ({\bf x-y})^2}{t^2} + 
\frac{\Delta t - 2t}{\Delta t} ~\frac{60 D^{b\ 2}_{0}}{t}\right)\right]  \nn \\
\ee
where
\be
P_0 ({\bf y}, t| {\bf x},0; D^b_{0})= \frac{e^{\frac{-{\bf x^2} }{4 t D^b_{0}}}}{2 \sqrt{t D^b_{0} \pi }}
\ee
is the unperturbed expression. 
%
To implement the renormalization, we choose to match the prediction of the two lowest  moments of this distribution, against the results of experiment or microscopic simulations at some time-scale $t^*$:
\be\label{R2ren}
\hspace{-7mm} \langle \Delta {\bf R}^2 (t^{\ast})\rangle_{exp} &\equiv& \langle \Delta \bar {\bf R}^2(t^{\ast})\rangle = 6 D_{ren} t^\ast,  \\ 
\label{R4ren}
\hspace{-7mm} \langle \Delta {\bf R}^4 (t^{\ast})\rangle_{exp} &\equiv&  \langle\Delta \bar {\bf R}^4(t^{\ast})\rangle = 60 D^2_{ren} t^{\ast 2} - C_{ren} t^\ast, 
\ee
where  $\Delta {\bf R}= ({\bf y} - {\bf x})$ and 
\be
\label{D_ren}
D_{ren}  &=& D_{0} \left[ 1 + 4 \xi^2 D_0 \Big( C^b_2 + Q_2 \right.   \\ 
        & & \hspace{8mm} \left. - \frac{20 D_0 }{\Delta t} \left( C^b_4 - Q_4\right) \Big) \right]+  o\left( \xi^4 \right)  ,  \nn \\
\label{C_ren}
C_{ren} &=& 1920 \xi^2 D_0^{4} \left( C^b_4 - Q_4\right)  +  o\left( \xi^4 \right) ,
\ee
are the renormalized constants, which are finite combinations of bare effective coefficients and counter-terms. Their numerical value is expected to run weakly with the matching time scale $t^\ast$. 

An important observation to make is that the mean-square displacement $\langle \Delta {\bf R}^2(t)\rangle$ retains its linear dependence on time $t$ (Einstein's law), even when quantum corrections are taken into account. 
In contrast, quantum corrections do affect the time dependence of  $\langle \Delta{\bf R}^4(t) \rangle$, by  introducing a linear term, which is absent in the classical diffusion limit.

Thus, the renormalized probability density including the leading-order quantum corrections reads:  
\be \label{RenProb}
&&\hspace{-3.5mm} \bar P({\bf y}, t| {\bf x},0) \simeq P_0({\bf y}, t| {\bf x},0; D_{ren}) \times \nn \\ 
&&\hspace{-5.5mm} \left[1 - C_{ren}~\left(\frac{({\bf y}-{\bf  x})^4}{t^3 D_{ren}} - 20\frac{({\bf y}-{\bf  x})^2}{t^2} + \frac{60 D_{ren}}{t}\right)\right].\nn\\
\ee

We empahsize that the $\xi^2$ expansion does not break down in the long-time limit. This can been seen directly from the expression (\ref{RenProb}), which shows that the perturbative corrections decay with time faster than the unperturbed term. In particular, the quantum excitation's dynamics reduces to the (unperturbed) classical over-damped diffusion, in the asymptotic long-time limit, implying that  the stochastic collisions contribute to quench the quantum effects.

\begin{table*}
\begin{tabular}{l|c|c|c|c|c|c}
$t^\ast$ [$ps$] &  1 & 5 & 7.5 & 10 & 12.5 & 15   \\
\hline 
$D_{ren} ~[\AA{}^2/ps]\times 10^2$  &  3,6 $\pm$ 0.2 &  3.0 $\pm$ 0.1 & 2.95 $\pm$ 0.05 & 2,90 $\pm$ 0.04 & 2.87 $\pm$ 0.04 & 2.85 $\pm$ 0.03 \\ 
$C_{ren} ~[\AA{}^4/ps]\times 10^6$  & -1.3 $\pm$ 0.1 & -2.7 $\pm$ 0.3 & -3.1 $\pm$ 0.3  & -3.2 $\pm$ 0.4  & -3.2 $\pm$ 0.4  & -3.1 $\pm$ 0.4
\end{tabular}
\caption{\label{DNA_ren_param} Renormalized coefficient in Eqs.~(\ref{D_ren}) and (\ref{C_ren}) fitted at different time scales $t^\ast$.}
\end{table*}
\begin{figure*}
 \includegraphics[width=8.7cm]{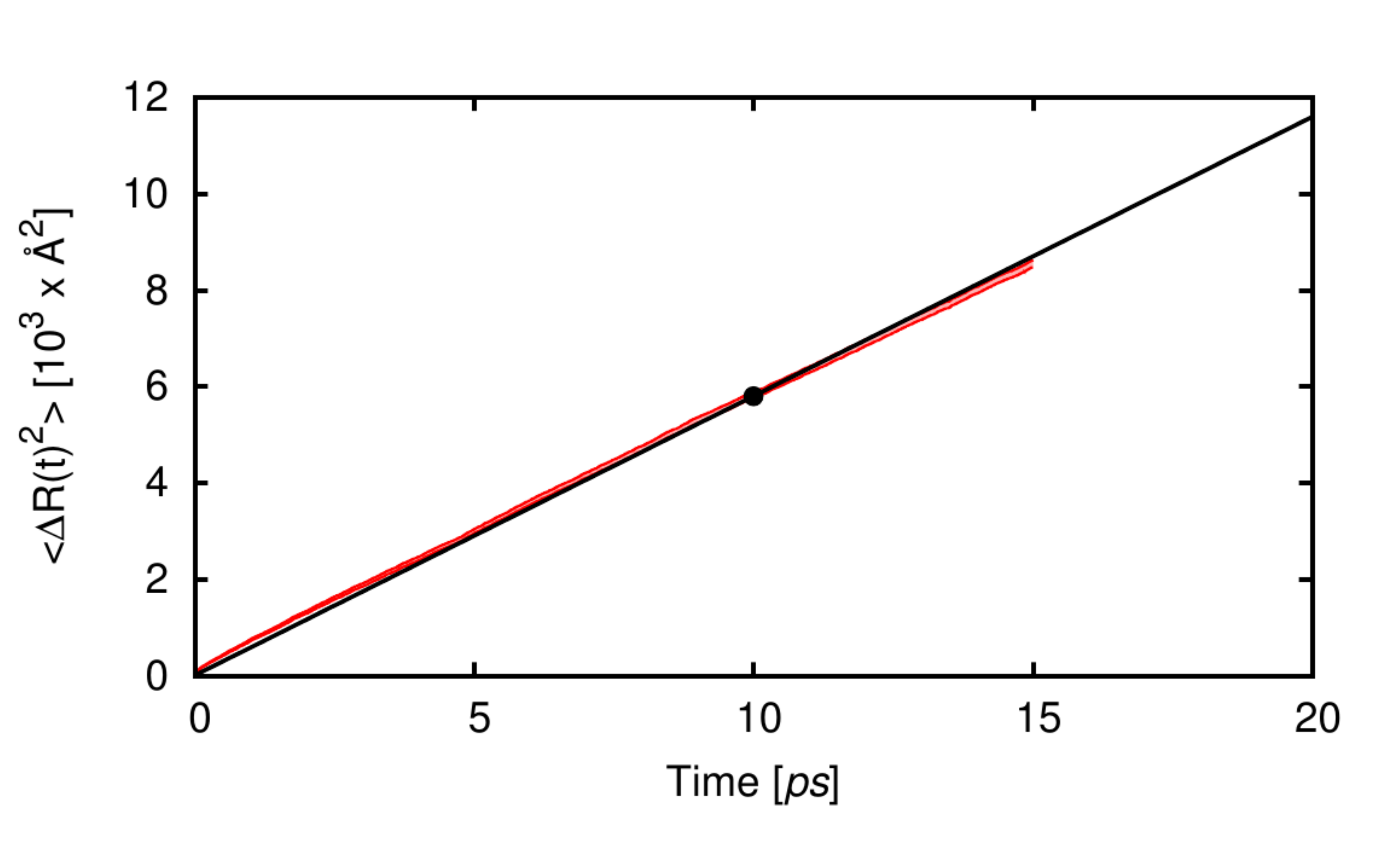}
 \includegraphics[width=8.7cm]{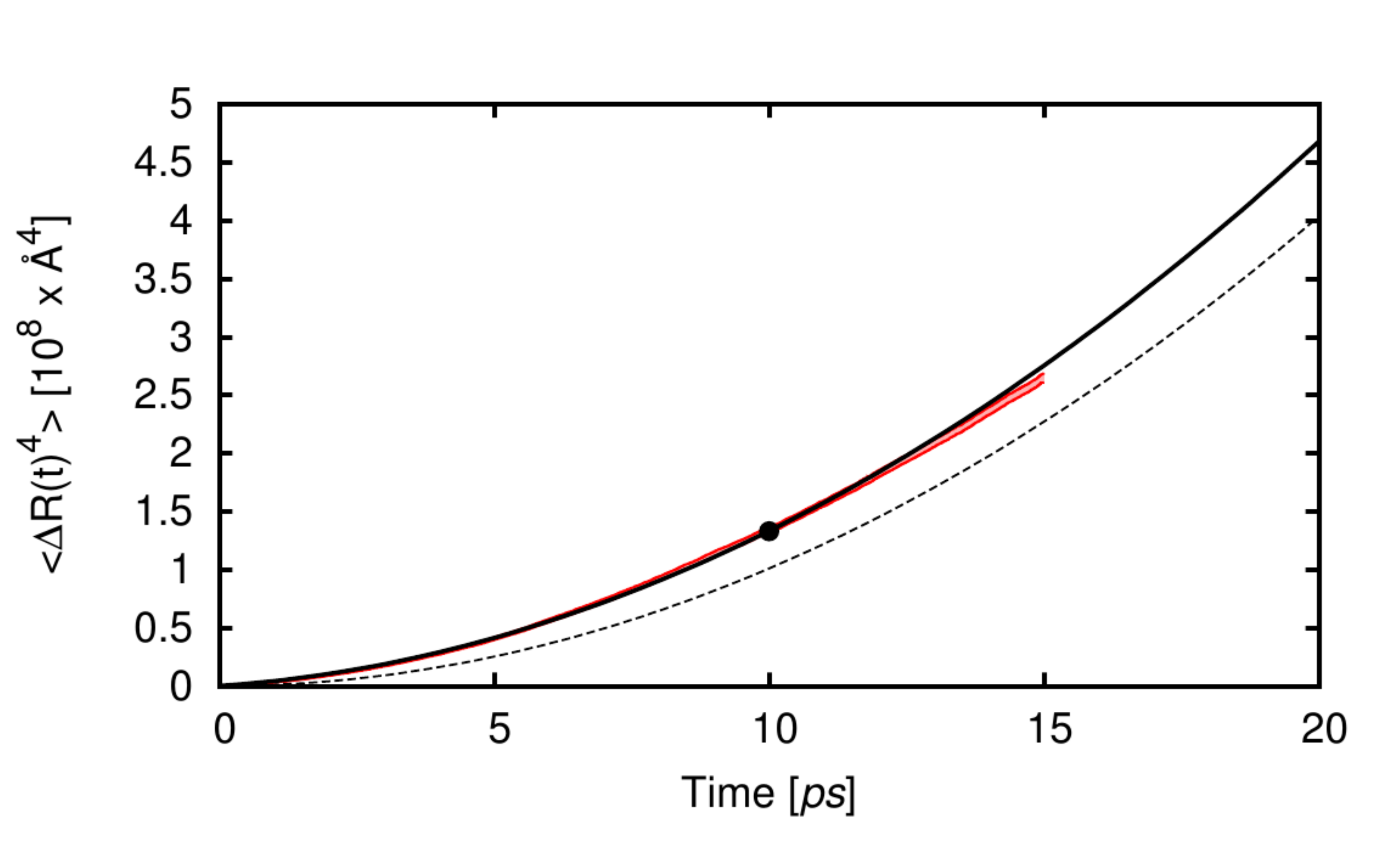}
 \caption{\label{plot} Time dependence of $\langle \Delta R^2(t) \rangle$ (left panel) and $\langle \Delta R^4(t) \rangle$ (right panel), in the microscopic model (red line) and in our effective theory (solid black line).  The dashed line in the right panel represents the prediction of a purely diffusive model (zero-th order contribution in $\xi^2$ expansion) and the black circles  represent the matching point (i.e. the renormalization time scale is $t^\ast=10$~ps). 
 The predictions in the microscopic model have been obtained in 200-base pair long molecule,  by averaging over 800 different trajectories generated using the algorithm defined in Ref.~\cite{boninsegna}. Beyond 15~fs these result become affected by fine-size effects.}
\end{figure*} 

\section{ Hole transport in a long homo-DNA molecular wire}
\label{application}
For illustration purposes, in this section we apply the effective theory developed above to investigate the dynamics of inelastic hole propagation along a  long homo-DNA molecule, which is regarded as an infinite molecular wire. To this end we first define a microscopic theory and then match the corresponding effective theory at a given time scale $t^*$ to define the renormalized parameters and finally use the effective theory to study the long-time and large-distance dynamics.  

We consider a very simple discrete model for the DNA conformational dynamics introduced in Ref.~\cite{PNASDNA}, in which the molecules vibration are effectively represented by the one-dimensional harmonic chain:
\be
V(Q)= \sum_{n=1}^N \frac{\kappa}{2} \left(  x_{i} -  x_{i-1} - a_0\right)^2 .
\ee
In this equation $x_{i}$ denotes the position of the $i-$th base pair, while $\kappa$=0.85~eV/\AA${}^2$ is the spring constant, while $a_0=3.4$ \AA{} is the equilibrium distance between two neighboring bases. In natural units (in which $\hbar=c=1$) the mass of each base pare is $M=2.44~10^{11}$~eV. 

The transfer integrals at the equilibrium position $f_{\bf l m}(Q_0)$ and its derivatives $f^a_{\bf l m}(Q_0)$, entering  Eq.~(\ref{hopp_param}), have been fixed in order to match the main features of the statistical distribution of transfer matrix elements for a homo-base DNA, computed microscopically in Ref.~\cite{DNA_theory3} from DFT-B electronic-structure calculations performed on snapshot of atomistic MD trajectories. 
Namely, we have set
\be
\hspace{-5mm} f^0_{\bf l m}   = & \langle t_{\bf l m}\rangle            & \equiv t_0 \left(\delta_{\bf l(m-1)} + \delta_{\bf l (m+1)}\right) - e_0 \delta_{\bf lm}, \\
\hspace{-5mm} f^{a}_{\bf l m} = & \sigma_{\bf l m} \sqrt{\beta \kappa } & \equiv t^{\prime} \left(\delta_{\bf l (m-1)} + \delta_{\bf l (m+1)}\right), 
\ee 
where $\langle t_{\bf l m}\rangle$ and $\sigma_{\bf l m}$ denote the average and the variance of the distribution  reported in Ref.~\cite{DNA_theory3}, leading to
$$
e_0=4.5~\text{eV}, \qquad t_0=0.03~\text{eV}, \qquad t^\prime = 0.15~\text{eV/\AA}.
$$
The system's temperature was set to $T=300 K$ and numerical simulations were performed on a 200-base pair molecule. 

We have studied the time-evolution of the probability density of an electronic hole,   initially prepared at the center of the molecule, using the algorithm introduced in Ref.\cite{boninsegna}, in which the stochastic conformational dynamics of the molecular wire is coupled to the quantum dynamics of the electronic hole. 

The formalism presented in the previous sections can be used to define a low-resolution perturbative effective theory for this molecular wire.  
In Fig.~\ref{plot}, we show the matching between the numerical simulations and analytic calculations in such an effective theory for the observables $\langle \Delta R^2(t) \rangle $ and $\langle \Delta R^4(t) \rangle$, fitted at the time scale $t^\ast=10$~ps (represented by a dot on the simulation curves). We note that the two approaches give consistent results. In particular, the inclusion of order $\xi^2$ corrections is necessary to reproduce the time-evolution of the $\langle R^4(t) \rangle$ moment. At times larger than $15$~ps finite size effects begin to affect the numerical simulations, and the microscopic model cannot be used to investigate the long distance propagation.

In Table~\ref{DNA_ren_param} we compare different values of the renormalized coefficient $D_{ren}$ and $C_{ren}$ corresponding to different renormalization time scales $t^\ast$. We observe that the effective parameters run only weakly with the renormalization scale $t^\ast$, as expected.

\section{Summary and Conclusions}
\label{conclusions}

In this work, we used the  EFT formalism to develop a rigorous effective description of the dissipative propagation of quantum excitations in macromolecular systems, in the long-time and large-distance regime. 

The underlying RG provides a rigorous framework to vary the degree of  space- and time- resolution, thereby allowing to access the meso-scopic regime. In particular, the spacial and time resolution powers of the EFT are set by the length cut-off scale $\lambda$ and the time scale $\Delta t$, respectively.

At low spatial resolution (i.e. for $\lambda$ much larger then the De Broglie's thermal wavelength $\lambda_B$) we have analytically computed the evolution of density of quantum excitations, by developing a perturbative expansion in the small parameter $\xi^2=\lambda_B^2/\lambda^2$. Our results show that, in the asymptotically long-time long-distance regime, the emerging dynamics of the quantum excitation  reduces to a classical diffusion process. At intermediate times, such a diffusive dynamics is modified by quantum corrections. 

We have illustrated our formalism by  applying it to study hole propagation in a long homo-DNA molecular wire. Comparison with numerical simulations show that, even at the leading-order level,  the effective theory yields very  accurate predictions. 

This framework can be applied to tackle a variety of problems involving quantum transport, ranging from charge transfer in biological or organic polymers, or crystalline organic transistors. Other potential applications involve the investigation of long-distance  exciton transfer  in natural or artificial photosynthetic complexes.

\acknowledgments
The authors are members of the Interdisciplinary Laboratory for Computational Science (LISC), a joint venture of Trento University and FBK foundation. The authors thank S. Taioli and S. a Beccara for a useful discussion.

\appendix
\section{Derivation of the $I$ and $J$ functionals in Eq. (\ref{Pfirst})}
\label{translation}
In section, \ref{effstoch} we have shown that the probability density for the quantum excitation at a given time $t$ can be written in the following form 
\be
\label{Pfirst2}
P({\bf y}, t| {\bf x},0) &=& \int \D {\bf X} \int \D {\bf Y} ~e^{\frac{i}{\hbar} \int_0^t d\tau \frac{1}{2}\,m_{ij}\,(\dot X_i \dot X_j-\dot Y_i \dot Y_j)} \nn\\
&&~\times e^{\frac{i}{\hbar} \left(I[{\bf X},{\bf Y}]+ J[{\bf X}, {\bf Y}]\right)}, 
\ee 
Where ${\bf X}[\tau]$ and ${\bf Y}[\tau]$ denote  paths in coordinate space of  quantum excitation described by the coherent 
fields $\phi^{'}$ and $\phi^{''}$. In this appendix, we explicitly derive the functionals $I[{\bf X},{\bf Y}]$ and $J[{\bf X},{\bf Y}]$, which originate from   translating into the first quantized formalism the field-theoretic functional $S_{int}$ appearing in Eq. (\ref{SintEFT}).

\subsection{$J[{\bf X},{\bf Y}]$ functional} 
The functional $J$ follows from the Hermitian  term:
\be
S_{J} &\equiv& \int dt'' \int dt' \int d\z \left\{\bar\Phi(\z,t')\Phi(\z, t') \bigg[A_v^0 \delta(t'-t'') \right. \nn\\
&& \left.   - i \hbar A_d^1 \frac{d}{d(t'-t'')}\bigg]~\bar\Phi(\z,t'')\tau_3\Phi(\z, t'')\right\}\nn\\
&=& \int \hspace{-2pt} dT \hspace{-2pt} \int d\tau \hspace{-2pt} \int \hspace{-2pt} d\z \left\{\bar\Phi(\z,T+\frac{\tau}{2})\Phi(\z,T+\frac{\tau}{2})\right. \Big(A_v^0 \delta(\tau) \nn\\
&& \left. - i \hbar A_v^1 \frac{d}{d\tau}\delta(\tau)\Big)\bar\Phi(\z,T-\frac{\tau}{2})\tau_3 \Phi(\z, T-\frac{\tau}{2})\right\}\nn\\
\ee
After expanding the $\Phi$ and $\bar \Phi$ fields into their components $\phi^{'}, \phi^{''}$, we obtain two symmetric terms:
$
S_J= S_{J_1} - S_{J_2}, 
$
where
\be
&&\hspace{-2mm} S_{J_1} = \int dT \int d\tau \int d\z \left\{\phi^{* '}(\z,T+\frac{\tau}{2})\phi^{'}(\z,T+\frac{\tau}{2})\right.\nn\\
&&\hspace{-2mm} \left. \left(A_v^0 \delta(\tau) - i \hbar A_v^1 \frac{d}{d\tau}\delta(\tau)\right)\phi^{'*}(\z,T-\frac{\tau}{2}) \phi^{'}(\z, T-\frac{\tau}{2})\right\}\nn\\
&&\hspace{-2mm} S_{J_2} = \int dT \int d\tau \int d\z \left\{\phi^{* ''}(\z,T+\frac{\tau}{2})\phi^{''}(\z,T+\frac{\tau}{2})\right.\nn\\
&&\hspace{-2mm} \left. \left(A_v^0 \delta(\tau) - i \hbar A_v^1 \frac{d}{d\tau}\delta(\tau)\right)\phi^{''*}(\z,T-\frac{\tau}{2}) \phi^{''}(\z, T-\frac{\tau}{2})\right\}\nn\\
\ee
 In first quantization representation, they translate into
\be
S_{J_1}&\to&J_1 = \int dT \int d\tau \int d\z \left\{\delta_\lambda(\z-{\bf X}(T+\frac{\tau}{2})) \right.\nn\\
&& \left. \left(A_v^0 \delta(\tau) - i \hbar A_v^1 \frac{d}{d\tau}\delta(\tau)\right)\delta_\lambda(\z-{\bf X}(T-\frac{\tau}{2})) )\right\}\nn\\
S_{J_2}&\to&J_2 = \int dT \int d\tau \int d\z \left\{\delta_\lambda(\z-{\bf Y}(T+\frac{\tau}{2})) \right.\nn\\
&& \left. \left(A_v^0 \delta(\tau) - i \hbar A_v^1 \frac{d}{d\tau}\delta(\tau)\right)\delta_\lambda(\z-{\bf Y}(T-\frac{\tau}{2})) )\right\}\nn\\ 
\ee
So that $S_{J}\to J= J_1 - J_2$. 

It is immediate to check that the terms proportional to $A_v^1$ vanish identically, while the terms proportional to $A_d^0$ cancel out in the difference between $J_1$ and $J_2$. Hence, $J=0$. 
This result is expected, indeed the interaction terms in the functional $S_J$ do not couple forward- and backward- propagating excitations, hence only contribute to the dressing of the one-body  propagator.

\subsection{$I[{\bf X},{\bf Y}]$ functional} 
The dissipative character of the effective theory comes from the non-Hermitian term in the functional (\ref{SintEFT}),
\be
 S_I &\equiv& \frac{i}{\beta^2 D \hbar} \int dt'' \int dt' \int d\z \left\{\bar\Phi(\z,t')\Phi(\z, t')\right.\nn\\
&&\hspace{-6.9mm} \left. \left[\left(A_d^0  - i \hbar A_d^1 \frac{d}{d(t'-t'')}\right)\delta(t'-t'')\right]\bar\Phi(\z,t'')\Phi(\z, t'')\right\}\nn\\
&=&\frac{i}{\beta^2 D \hbar} \int dT \int d\tau \int d\z \left\{\bar\Phi(\z,T+\frac{\tau}{2})\Phi(\z,T+\frac{\tau}{2})\right.\nn\\
&& \hspace{-3.7mm}\left. \left(A_d^0 \delta(\tau) - i  \hbar A_d^1 \frac{d}{d\tau}\delta(\tau)\right)\bar\Phi(\z,T-\frac{\tau}{2})\Phi(\z, T-\frac{\tau}{2})\right\}\nn\\
\ee
After expanding the $\Phi$ and $\bar \Phi$ fields into their components we obtain:
\be
S_I= S_{I_1} + S_{I_2} + S_{I_3},
\ee
where
\be
S_{I_1} &=& \frac{i}{\beta^2 D \hbar} \int \hspace{-1mm}dT \hspace{-1.5mm}\int \hspace{-1mm}d\tau \hspace{-1.5mm}\int \hspace{-1mm}d\z \bigg\{\phi^{' *}(\z,T+\frac{\tau}{2})\phi^{'}(\z, T+\frac{\tau}{2})\nn\\
&& \hspace{-0.8cm}\left. \left(A_d^0 \delta(\tau) - i \hbar A_d^1 \frac{d}{d\tau}\delta(\tau)\right)\phi^{'*}(\z,T-\frac{\tau}{2})\phi^{'}(\z, T-\frac{\tau}{2})\right\}\nn\\
S_{I_2} &=& \frac{i}{\beta^2 D \hbar} \int \hspace{-1mm}dT \hspace{-1.5mm}\int \hspace{-1mm}d\tau \hspace{-1.5mm}\int \hspace{-1mm}d\z \left\{\phi^{'' *}(\z,T+\frac{\tau}{2})\phi^{''}(\z,T+\frac{\tau}{2})\right.\nn\\
&& \hspace{-0.9cm} \left. \left(A_d^0 \delta(\tau) - i \hbar A_d^1 \frac{d}{d\tau}\delta(\tau)\right)\phi^{''*}(\z,T-\frac{\tau}{2})\phi^{''}(\z, T-\frac{\tau}{2})\right\}\nn\\
S_{I_3} &=& \frac{-2 i }{\beta^2 D \hbar} \int \hspace{-1mm}dT \hspace{-1.5mm}\int \hspace{-1mm}d\tau \hspace{-1.5mm}\int \hspace{-1mm}d\z \left\{\phi^{' *}(\z,T+\frac{\tau}{2})\phi^{'}(\z, T+\frac{\tau}{2})\right.\nn\\
&& \hspace{-0.9cm} \left. \left(A_d^0 \delta(\tau) - i \hbar A_d^1 \frac{d}{d\tau}\delta(\tau)\right)\phi^{''*}(\z,T-\frac{\tau}{2})\phi^{''}(\z, T-\frac{\tau}{2})\right\}\nn\\
\ee
Let's begin by analyzing the $S_{I_1}$ part. In first quantization  representation it translates as
\be
S_{I_1}&\to &I_1= \frac{i}{\beta^2 D \hbar} \int_0^t dT \hspace{-1mm}\int d\z \, A_d^0\delta_\lambda({\bf X}-\z) ~\delta_\lambda({\bf X}-\z)\nn\\
&&+ i  \hbar A_d^1  \int d\tau \int d\z  
\delta(\tau)\,\frac{d}{d\tau} \left[~\delta_\lambda(\z- {\bf X}(T+\frac{\tau}{2}))\right. \nn\\
&& \left. ~\delta_\lambda(\z- {\bf X}( T-\frac{\tau}{2}))\right]
\ee
The term proportional to $A_d^1$ vanishes identically, while the term proportional to $A_d^0$ is independent on the paths and reads (setting to $d=3$ the number of spatial dimensions)
\be
I_1 = \frac{i t}{\beta^2 D \hbar} \frac{\sqrt{\det m}}{ (4 \text{Tr}\,m~\lambda^2 \pi)^{3/2}}~A_d^0
\ee
Clearly, by symmetry, we find that $S_{I_2}\to I_2= I_1$. 

Let us now consider the cross-term $S_{I_3}$, which couples forward and backward propagating paths: 
\be
&&\hspace{-0.5cm}S_{I_3}= \frac{-2 i A_d^0}{\beta^2 D \hbar} \int\hspace{-1mm} dT \hspace{-1mm} \int \hspace{-1mm}d\z \,\phi^{' \ast}(\z, T)\phi^{'}(\z, T) \phi^{'' \ast}(\z, T)\phi^{''}(\z, T)\nn\\
&&\hspace{1.6mm}+ \frac{ 2 A_d^1 }{\beta^2 D} \int \hspace{-1mm}dT \hspace{-1mm}\int \hspace{-1mm}d\tau \hspace{-1mm}\int \hspace{-1mm}d\z ~ \delta(\tau)\,\frac{d}{d\tau} \Big[\phi^{' *}(\z,T+\frac{\tau}{2})\phi^{'}(\z,T+\frac{\tau}{2}) \nn\\
&&\hspace{20mm} \left. \phi^{''*}(\z,T-\frac{\tau}{2})\phi^{''}(\z, T-\frac{\tau}{2})\right].
\ee
Translating into the first quantization form, we have $S_{I_3}\to I_3$, with 
\be\mbox{}
&&I_3= \frac{-2 i A_d^0}{\beta^2 D \hbar} \int dT   \int d\z ~\delta_\lambda(\z- {\bf X}(T)) \delta_\lambda(\z- {\bf Y}(T)) \nn\\
&&+ \frac{ 2 A_d^1 }{\beta^2 D} \int dT \int d\tau \int d\z  
\delta(\tau)\,\frac{d}{d\tau} \left[~\delta_\lambda(\z- {\bf X}(T+\frac{\tau}{2}))\right. \nn\\
&& \left. ~\delta_\lambda(\z- {\bf Y}( T-\frac{\tau}{2}))\right]
\ee  
After writing explicitly the smeared representation of the $\delta$-function and evaluating the corresponding  Gaussian integrals we find:
\be
&&I_3= \frac{\sqrt{\det{m}} }{\beta^2 D \hbar\, ( 4\,\text{Tr}\,m~\pi \lambda^2)^{3/2}} \int_0^t dT\left\{ e^{\frac{-m_{ij}}{4 \text{Tr} m\, \lambda^2}(X- Y)_i (X- Y)_j}\right. \nn\\
&& \left. \left(-i 2 A_d^0  + \frac{ \hbar A_d^1}{2 \text{Tr}\,m~\lambda^2}~ m_{ij}~( Y- X)_i (\dot Y+ \dot X)_j\right) \right\}
\ee

\newpage

\begin{thebibliography}{}

\expandafter\ifx\csname natexlab\endcsname\relax\def\natexlab#1{#1}\fi
\expandafter\ifx\csname bibnamefont\endcsname\relax
  \def\bibnamefont#1{#1}\fi
\expandafter\ifx\csname bibfnamefont\endcsname\relax
  \def\bibfnamefont#1{#1}\fi
\expandafter\ifx\csname citenamefont\endcsname\relax
  \def\citenamefont#1{#1}\fi
\expandafter\ifx\csname url\endcsname\relax
  \def\url#1{\texttt{#1}}\fi
\expandafter\ifx\csname urlprefix\endcsname\relax\def\urlprefix{URL }\fi
\providecommand{\bibinfo}[2]{#2}
\providecommand{\eprint}[2][]{\url{#2}}


\bibitem[{\citenamefont{Engel et~al.}(2007)\citenamefont{Engel, Calhoun, Read,
  Ahn, Mancal, Cheng, Blankenship, and Fleming}}]{photo_exp1}
\bibinfo{author}{\bibfnamefont{G.~S.} \bibnamefont{Engel}},
  \bibinfo{author}{\bibfnamefont{T.~R.} \bibnamefont{Calhoun}},
  \bibinfo{author}{\bibfnamefont{E.~L.} \bibnamefont{Read}},
  \bibinfo{author}{\bibfnamefont{T.-K.} \bibnamefont{Ahn}},
  \bibinfo{author}{\bibfnamefont{T.}~\bibnamefont{Mancal}},
  \bibinfo{author}{\bibfnamefont{Y.-C.} \bibnamefont{Cheng}},
  \bibinfo{author}{\bibfnamefont{R.~E.} \bibnamefont{Blankenship}},
  \bibnamefont{and} \bibinfo{author}{\bibfnamefont{G.~R.}
  \bibnamefont{Fleming}}, \bibinfo{journal}{Nature}
  \textbf{\bibinfo{volume}{446}}, \bibinfo{pages}{782} (\bibinfo{year}{2007}).
  
\bibitem{nano-bio-electronics} \emph{Nano- Bio-Electronic, Photonic and MEMS Packaging},
 edited by C. P. Wong, Kyoung-Sik Moon and Yi Li, Springer (New York, Dordrecht, Heidelberg, London), 2010. 
 
 \bibitem[{\citenamefont{Frederiksen et~al.}(2007)\citenamefont{Frederiksen,
  Paulsson, Brandbyge, and Jauho}}]{NEGF1}
\bibinfo{author}{\bibfnamefont{T.}~\bibnamefont{Frederiksen}},
  \bibinfo{author}{\bibfnamefont{M.}~\bibnamefont{Paulsson}},
  \bibinfo{author}{\bibfnamefont{M.}~\bibnamefont{Brandbyge}},
  \bibnamefont{and} \bibinfo{author}{\bibfnamefont{A.-P.} \bibnamefont{Jauho}},
  \bibinfo{journal}{Phys. Rev. B} \textbf{\bibinfo{volume}{75}},
  \bibinfo{pages}{205413} (\bibinfo{year}{2007}).

\bibitem[{\citenamefont{James and Tour}(2005)}]{organic1}
\bibinfo{author}{\bibfnamefont{D.}~\bibnamefont{James}} \bibnamefont{and}
  \bibinfo{author}{\bibfnamefont{J.}~\bibnamefont{Tour}},
  \bibinfo{journal}{Top. Curr. Chem.} \textbf{\bibinfo{volume}{257}},
  \bibinfo{pages}{33} (\bibinfo{year}{2005}).

\bibitem[{\citenamefont{Dimitrakopoulos and Malenfant}(2002)}]{organic2}
\bibinfo{author}{\bibfnamefont{C.~D.} \bibnamefont{Dimitrakopoulos}}
  \bibnamefont{and} \bibinfo{author}{\bibfnamefont{P.~R.~L.}
  \bibnamefont{Malenfant}}, \bibinfo{journal}{Adv. Mater.}
  \textbf{\bibinfo{volume}{14}}, \bibinfo{pages}{99} (\bibinfo{year}{2002}).

\bibitem[{\citenamefont{Torsi et~al.}(2001)\citenamefont{Torsi, Cioffi,
  Di~Franco, Sabbatini, Zambonin, and Bleve-Zacheo}}]{organic3}
\bibinfo{author}{\bibfnamefont{L.}~\bibnamefont{Torsi}},
  \bibinfo{author}{\bibfnamefont{N.}~\bibnamefont{Cioffi}},
  \bibinfo{author}{\bibfnamefont{C.}~\bibnamefont{Di~Franco}},
  \bibinfo{author}{\bibfnamefont{L.}~\bibnamefont{Sabbatini}},
  \bibinfo{author}{\bibfnamefont{P.~G.} \bibnamefont{Zambonin}},
  \bibnamefont{and}
  \bibinfo{author}{\bibfnamefont{T.}~\bibnamefont{Bleve-Zacheo}},
  \bibinfo{journal}{Solid-State Electr.} \textbf{\bibinfo{volume}{45}},
  \bibinfo{pages}{1479} (\bibinfo{year}{2001}).

\bibitem[{\citenamefont{Cheung et~al.}(2009)\citenamefont{Cheung, McMahon, and
  Troisi}}]{P3HT1}
\bibinfo{author}{\bibfnamefont{D.}~\bibnamefont{Cheung}},
  \bibinfo{author}{\bibfnamefont{D.~P.} \bibnamefont{McMahon}},
  \bibnamefont{and} \bibinfo{author}{\bibfnamefont{A.}~\bibnamefont{Troisi}},
  \bibinfo{journal}{J. Phys. Chem. B} \textbf{\bibinfo{volume}{113}},
  \bibinfo{pages}{9393} (\bibinfo{year}{2009}).

\bibitem[{\citenamefont{Mallajosyula and Pati}(2010)}]{DNAreview}
\bibinfo{author}{\bibfnamefont{S.~S.} \bibnamefont{Mallajosyula}}
  \bibnamefont{and} \bibinfo{author}{\bibfnamefont{S.~K.} \bibnamefont{Pati}},
  \bibinfo{journal}{J. Phys. Chem. Lett.} \textbf{\bibinfo{volume}{1}},
  \bibinfo{pages}{1881} (\bibinfo{year}{2010}).

\bibitem[{\citenamefont{Guti\'errez et~al.}(2009)\citenamefont{Guti\'errez,
  Caetano, Woiczikowski, Kubar, Elstner, and Cuniberti}}]{DNA_theory1}
\bibinfo{author}{\bibfnamefont{R.}~\bibnamefont{Guti\'errez}},
  \bibinfo{author}{\bibfnamefont{R.~A.} \bibnamefont{Caetano}},
  \bibinfo{author}{\bibfnamefont{B.~P.} \bibnamefont{Woiczikowski}},
  \bibinfo{author}{\bibfnamefont{T.}~\bibnamefont{Kubar}},
  \bibinfo{author}{\bibfnamefont{M.}~\bibnamefont{Elstner}}, \bibnamefont{and}
  \bibinfo{author}{\bibfnamefont{G.}~\bibnamefont{Cuniberti}},
  \bibinfo{journal}{Phys. Rev. Lett.} \textbf{\bibinfo{volume}{102}},
  \bibinfo{pages}{208102} (\bibinfo{year}{2009}).
  
  \bibitem[{\citenamefont{Kubar et~al.}(2008)\citenamefont{Kubar, Woiczikowski,
  Cuniberti, and Elstner}}]{DNA_theory2}
\bibinfo{author}{\bibfnamefont{T.}~\bibnamefont{Kubar}},
  \bibinfo{author}{\bibfnamefont{P.}~\bibnamefont{Woiczikowski}},
  \bibinfo{author}{\bibfnamefont{G.}~\bibnamefont{Cuniberti}},
  \bibnamefont{and} \bibinfo{author}{\bibfnamefont{M.}~\bibnamefont{Elstner}},
  \bibinfo{journal}{J. Phys. Chem. B} \textbf{\bibinfo{volume}{112}},
  \bibinfo{pages}{7937} (\bibinfo{year}{2008}).

  \bibitem[{\citenamefont{Woiczikowski et~al.}(2009)\citenamefont{Woiczikowski,
  Kubar, Guti\'{e}rrez, Caetano, Cuniberti, and Elstner}}]{DNA_theory3}
\bibinfo{author}{\bibfnamefont{P.~B.} \bibnamefont{Woiczikowski}},
  \bibinfo{author}{\bibfnamefont{T.}~\bibnamefont{Kubar}},
  \bibinfo{author}{\bibfnamefont{R.}~\bibnamefont{Guti\'{e}rrez}},
  \bibinfo{author}{\bibfnamefont{R.~A.} \bibnamefont{Caetano}},
  \bibinfo{author}{\bibfnamefont{G.}~\bibnamefont{Cuniberti}},
  \bibnamefont{and} \bibinfo{author}{\bibfnamefont{M.}~\bibnamefont{Elstner}},
  \bibinfo{journal}{J. Chem. Phys.} \textbf{\bibinfo{volume}{130}},
  \bibinfo{pages}{215104} (\bibinfo{year}{2009}).


\bibitem[{\citenamefont{Chen et~al.}(2012)\citenamefont{Chen, Hong, and
  Huang}}]{peptide}
\bibinfo{author}{\bibfnamefont{Y.-S.} \bibnamefont{Chen}},
  \bibinfo{author}{\bibfnamefont{M.-Y.} \bibnamefont{Hong}}, \bibnamefont{and}
  \bibinfo{author}{\bibfnamefont{G.~S.} \bibnamefont{Huang}},
  \bibinfo{journal}{Nature Nanotech.} \textbf{\bibinfo{volume}{7}},
  \bibinfo{pages}{197} (\bibinfo{year}{2012}).


\bibitem[{\citenamefont{Mohseni et~al.}(2008)\citenamefont{Mohseni, Rebentrost,
  Lloyd, and Aspuru-Guzik}}]{photo_theory1}
\bibinfo{author}{\bibfnamefont{M.}~\bibnamefont{Mohseni}},
  \bibinfo{author}{\bibfnamefont{P.}~\bibnamefont{Rebentrost}},
  \bibinfo{author}{\bibfnamefont{S.}~\bibnamefont{Lloyd}}, \bibnamefont{and}
  \bibinfo{author}{\bibfnamefont{A.}~\bibnamefont{Aspuru-Guzik}},
  \bibinfo{journal}{J. Chem. Phys.} \textbf{\bibinfo{volume}{129}},
  \bibinfo{pages}{174106} (\bibinfo{year}{2008}).

\bibitem[{\citenamefont{Rebentrost
  et~al.}(2009{\natexlab{a}})\citenamefont{Rebentrost, Mohseni, Kassal, Llyod,
  and Aspuru-Guzik}}]{photo_theory2}
\bibinfo{author}{\bibfnamefont{P.}~\bibnamefont{Rebentrost}},
  \bibinfo{author}{\bibfnamefont{M.}~\bibnamefont{Mohseni}},
  \bibinfo{author}{\bibfnamefont{I.}~\bibnamefont{Kassal}},
  \bibinfo{author}{\bibfnamefont{S.}~\bibnamefont{Llyod}}, \bibnamefont{and}
  \bibinfo{author}{\bibfnamefont{A.}~\bibnamefont{Aspuru-Guzik}},
  \bibinfo{journal}{New J. Phys.} \textbf{\bibinfo{volume}{11}},
  \bibinfo{pages}{033003} (\bibinfo{year}{2009}{\natexlab{a}}).

\bibitem[{\citenamefont{Rebentrost
  et~al.}(2009{\natexlab{b}})\citenamefont{Rebentrost, Mohseni, and
  Aspuru-Guzik}}]{photo_theory3}
\bibinfo{author}{\bibfnamefont{P.}~\bibnamefont{Rebentrost}},
  \bibinfo{author}{\bibfnamefont{M.}~\bibnamefont{Mohseni}}, \bibnamefont{and}
  \bibinfo{author}{\bibfnamefont{A.}~\bibnamefont{Aspuru-Guzik}},
  \bibinfo{journal}{J. Phys. Chem. B} \textbf{\bibinfo{volume}{113}},
  \bibinfo{pages}{9942} (\bibinfo{year}{2009}{\natexlab{b}}).
\bibitem{plenio2} M.B. Plenio and S.F. Huelga, New J. Phys. {\bf 10}, 113019 (2008).
\bibitem{plenio3} F. Caruso, A.W. Chin, A. Datta, S.F. Huelga and M.B. Plenio. J. Chem. Phys. {\bf 131}, 105106 (2009).

\bibitem[{\citenamefont{Chin et~al.}(2013)\citenamefont{Chin, Prior, Rosenbach,
  Caycedo-Soler, Huelga, and Plenio}}]{photo_theory4}
\bibinfo{author}{\bibfnamefont{A.~W.} \bibnamefont{Chin}},
  \bibinfo{author}{\bibfnamefont{J.}~\bibnamefont{Prior}},
  \bibinfo{author}{\bibfnamefont{R.}~\bibnamefont{Rosenbach}},
  \bibinfo{author}{\bibfnamefont{F.}~\bibnamefont{Caycedo-Soler}},
  \bibinfo{author}{\bibfnamefont{S.~F.} \bibnamefont{Huelga}},
  \bibnamefont{and} \bibinfo{author}{\bibfnamefont{M.~B.}
  \bibnamefont{Plenio}}, \bibinfo{journal}{Nature Phys.}
  \textbf{\bibinfo{volume}{9}}, \bibinfo{pages}{113} (\bibinfo{year}{2013}).
 
 
\bibitem{open_quantum_system_book} H.-P. Breuer and F. Petruccione, \emph{The Theory of Open Quantum Systems}, 2nd ed. (Pearson Education Limited, Harlow, 2001). 



\bibitem[{\citenamefont{Boninsegna and Faccioli}(2012)}]{boninsegna}
\bibinfo{author}{\bibfnamefont{L.}~\bibnamefont{Boninsegna}} \bibnamefont{and}
  \bibinfo{author}{\bibfnamefont{P.}~\bibnamefont{Faccioli}},
  \bibinfo{journal}{J. Chem. Phys.} \textbf{\bibinfo{volume}{136}},
  \bibinfo{pages}{214111} (\bibinfo{year}{2012}).


\bibitem{Elia1} 
\bibinfo{author}{\bibfnamefont{E.}~\bibnamefont{Schneider}},
\bibinfo{author}{\bibfnamefont{S.}~\bibnamefont{a Beccara}}, \bibnamefont{and} 
\bibinfo{author}{\bibfnamefont{P.}~\bibnamefont{Faccioli}},
\bibinfo{journal}{Phys.\ Rev.\ B} \textbf{\bibinfo{volume}{88}},
 \bibinfo{pages}{08542} (\bibinfo{year}{2013}). 

\bibitem[{\citenamefont{Grabert et~al.}(1988)\citenamefont{Grabert, Schramm,
  and Ingold}}]{PIreview}
\bibinfo{author}{\bibfnamefont{H.}~\bibnamefont{Grabert}},
  \bibinfo{author}{\bibfnamefont{P.}~\bibnamefont{Schramm}}, \bibnamefont{and}
  \bibinfo{author}{\bibfnamefont{G.-L.} \bibnamefont{Ingold}},
  \bibinfo{journal}{Phys. Rep.} \textbf{\bibinfo{volume}{168}},
  \bibinfo{pages}{115} (\bibinfo{year}{1988}).


\bibitem{EFT1} P. Lepage, arXiv:nucl-th/9706029 (unpublished). 
\bibitem{EFT2} A. Manohar, arXiv:hep-ph/9606222 (unpublished)


\bibitem{graphene1} P. Faccioli and E. Lipparini, Phys. Rev. B {\bf 80}, 045405 (2009)
\bibitem{graphene2} M. Bazzanella, P. Faccioli and E. Lipparini, Phys. Rev. B {\bf 82}, 205422 (2010) 

\bibitem{RGMD1} O. Corradini, P. Faccioli and H. Orland, Phys. Rev. E {\bf 80}, 061112 (2009).
\bibitem{RGMD2} P. Faccioli, J. Chem. Phys. {\bf 133}, 164106 (2010).
\bibitem{RGMD3} T. Ichinomiya, J. Comp.  Phys. {\bf 251}, 319 (2013). 

\bibitem[{\citenamefont{Caldeira and Leggett}(1981)}]{leggett-caldeira}
\bibinfo{author}{\bibfnamefont{A.~O.} \bibnamefont{Caldeira}} \bibnamefont{and}
  \bibinfo{author}{\bibfnamefont{A.~J.} \bibnamefont{Leggett}},
  \bibinfo{journal}{Phys. Rev. Lett.} \textbf{\bibinfo{volume}{46}},
  \bibinfo{pages}{211} (\bibinfo{year}{1981}).
\bibitem{LeBellac} M. Le Bellac, \emph{Quantum and Statistical Field Theory}, (Oxford University Press Inc., New York, 1991).
\bibitem{OMfunctional} L. Onsager and S. Machlup, Phys. Rev. {\bf 91}, 1505 (1953).

\bibitem{measuring-mobility} J. U. Wallace, \emph{``Carrier Mobility in Organic Charge Transport Materials: Methods of Measurement, Analysis, and Modulation"}, PhD thesis, Department of Chemical Engineering of the University of Rochester (2009) (unpublished).

\bibitem{Troisi} A. Troisi,  J. Chem. Phys. {\bf 134}, 034702 (2011)

\bibitem{QL1} L. Machura, M. Kostur, P. H\"anggi, P. Talkner and J. Luczka, Phys. Rev. {\bf E 70}, 031107 (2004). 
\bibitem{QL2}  W. T. Coffey, Y. P. Kalmykov, S. V. Titov, and B. P. Mulligan, J. Phys. {\bf A 40}, F91(2007). 
\bibitem{QL3} W. T. Coffey, Y. P. Kalmykov, S. V. Titov, L. Cleary, Phys. Rev. {\bf E 78} 031114 (2008).   
\bibitem{QL4} S. a Beccara, G. Garberoglio, and P. Faccioli, J. Chem. Phys. {\bf 135}, 034103 (2011).

\bibitem{PNASDNA} E. Conwell, S. Rakhmanova, Proc. Natl. Acad. Sci. USA 97, 4556 (2000)
\end{thebibliography}
\end{document}